\documentclass[acmlarge]{acmart}
\usepackage{graphicx}
\usepackage{subfigure}
 \usepackage{algorithm}
\usepackage{algorithmic}
\usepackage{booktabs} 
\usepackage{amsmath}

 \usepackage{multirow}
 \usepackage{rotating}
 \usepackage{amssymb}
 \usepackage{bbding}
 \usepackage{amsthm}
 \usepackage{bm}
 \usepackage{epstopdf}
 \usepackage{enumerate}
 \usepackage{array}
\usepackage{hyperref}

\newcommand{\Lapl}{\mathbf{\mathop{\mathcal{L}}}}

\newcommand{\Space}[1]{\mathbb{#1}}
\newcommand{\Set}[1]{\mathcal{#1}}

\newcommand{\ie}{\emph{i.e., }}
\newcommand{\eg}{\emph{e.g., }}
\newcommand{\etal}{\emph{et al.}}

\newcommand{\wrt}{\emph{w.r.t. }}
\newcommand{\cf}{\emph{cf. }}
\newcommand{\aka}{\emph{aka. }}

\AtBeginDocument{%
  \providecommand\BibTeX{{%
    \normalfont B\kern-0.5em{\scshape i\kern-0.25em b}\kern-0.8em\TeX}}}

\setcopyright{acmcopyright}
\copyrightyear{2020}
\acmYear{2020}
\acmDOI{10.1145/1122445.1122456}

\begin{document}

\title{Bias and Debias in Recommender System: A Survey and Future Directions}

\author{Jiawei Chen}
\email{	cjwustc@ustc.edu.cn}
\author{Hande Dong}
\email{donghd@mail.ustc.edu.cn}
\affiliation{%
  \institution{University of Science and Technology of China}
  \city{Hefei}
  \country{China}
  }

\author{Xiang Wang}
\email{	 xiangwang@u.nus.edu}
\author{Fuli Feng}
\email{fulifeng93@gmail.com}
\affiliation{%
  \institution{National University of Singapore}
  \city{Singapore}
  \country{Singapore}
  }

\author{Meng Wang}
\email{eric.mengwang@gmail.com}
\affiliation{%
  \institution{Hefei University of Technology}
  \city{Hefei}
  \country{China}
  }

\author{Xiangnan He$^\dagger$}
\thanks{$^\dagger$Corresponding author: hexn@ustc.edu.cn}
\email{hexn@ustc.edu.cn}
\affiliation{%
  \institution{University of Science and Technology of China}
  \city{Hefei}
  \country{China}
  }

\renewcommand{\shortauthors}{Chen and Dong, et al.}

\begin{abstract}
While recent years have witnessed a rapid growth of research papers on recommender system (RS), most of the papers focus on inventing machine learning models to better fit user behavior data. However, user behavior data
is observational rather than experimental. This makes various biases widely exist in the data, including but not limited to selection bias, position bias, exposure bias, and popularity bias. Blindly fitting the data without considering the inherent biases will result in many serious issues, e.g., the discrepancy between offline evaluation and online metrics, hurting user satisfaction and trust on the recommendation service, etc. To transform the large volume of research models into practical improvements, it is highly urgent to explore the impacts of the biases and perform debiasing when necessary.
When reviewing the papers that consider biases in RS, we find that, to our surprise, the studies are rather fragmented and lack a systematic organization. The terminology ``bias'' is widely used in the literature, but its definition is usually vague and even inconsistent across papers. This motivates us to provide a systematic survey of existing work on RS biases. In this paper, we first summarize seven types of biases in recommendation, along with their definitions and characteristics. We then provide a taxonomy to position and organize the existing work on recommendation debiasing. Finally, we identify some open challenges and envision some future directions, with the hope of inspiring more research work on this important yet less investigated topic. The summary of debiasing methods reviewed in this survey can be found at \url{https://github.com/jiawei-chen/RecDebiasing}.
\end{abstract}



\maketitle

\section{Introduction}
 \allowdisplaybreaks[4]
Being able to provide personalized suggestions to each user, recommender system (RS) has been recognized as the most effective way to alleviate information overloading. It not only facilitates users seeking information, but also benefits content providers with more potentials of making profits. Nowadays, recommendation techniques have been intensively used in countless applications, e.g., E-commerce platforms (Alibaba, Amazon), social networks (Facebook, Weibo), video-sharing platforms (YouTube, TikTok), lifestyle apps (Yelp, Meituan), and so on. As such, the importance of RS cannot be overstated especially in the era that the information overload issue becomes increasingly serious.

\textbf{Ubiquity of Biases in RS.} Although RS has generated large impacts in a wide range of applications, it faces many bias problems which are challenging to handle and may deteriorate the recommendation effectiveness.
Bias is common in RS for the following factors. (1) User behavior data, which lays the foundation for recommendation model training, is observational rather than experimental.
The main reason is that a user generates behaviors on the basis of the exposed items, making the observational data confounded by the exposure mechanism of the system and the self-selection of the user.
(2) Items are not evenly presented in the data, e.g., some items are more popular than others and thus receive more user behaviors. As a result, these popular items would have a larger impact on the model training, making the recommendations biased towards them. The same situation applies to the user side.
(3) One nature of RS is the feedback loop --- the exposure mechanism of the RS determines user behaviors, which are circled back as the training data for the RS. Such feedback loop not only creates biases but also intensifies biases over time, resulting in ``the rich get richer'' Matthew effect.

\begin{figure}[t]
  \centering
\subfigure{
\includegraphics[width=0.5\textwidth]{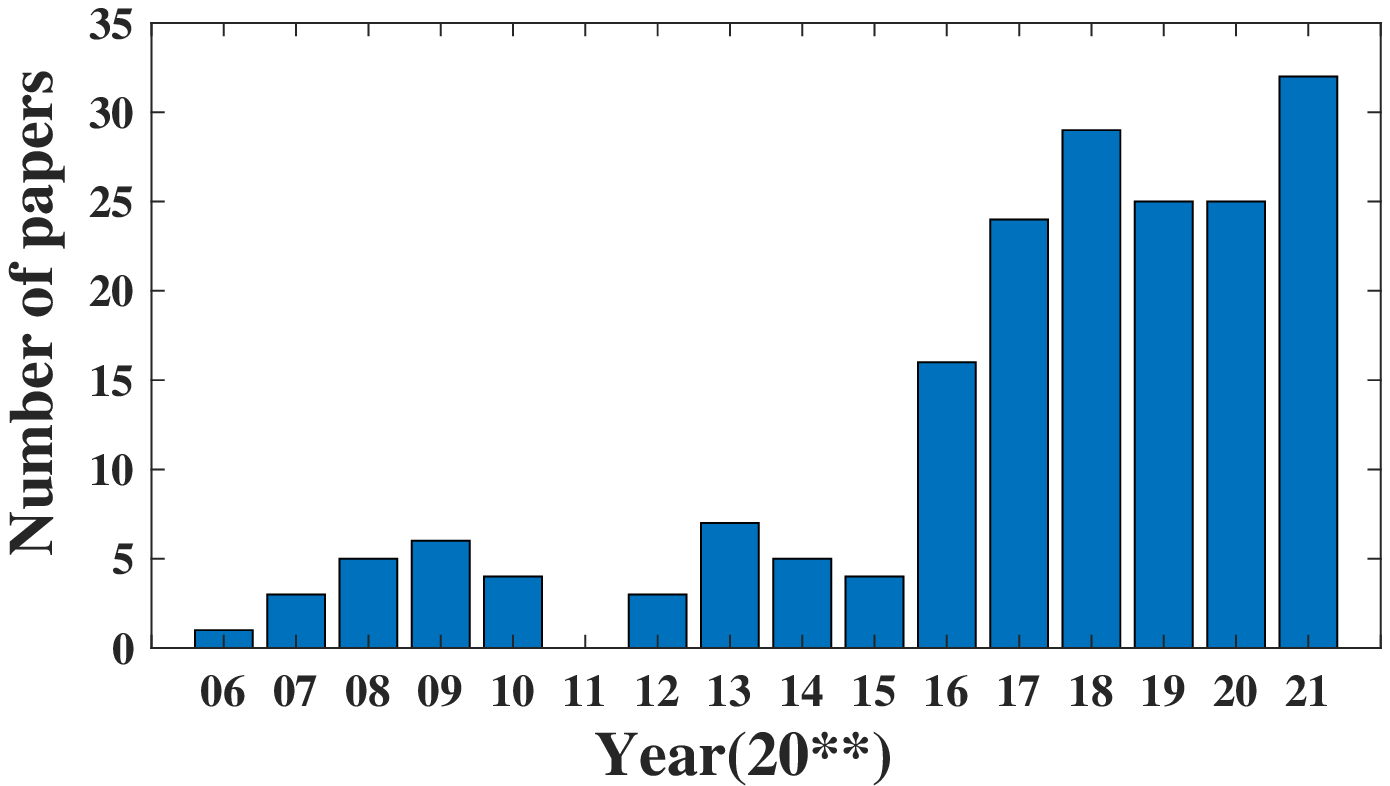}
}
\subfigure{
\includegraphics[width=0.45\textwidth]{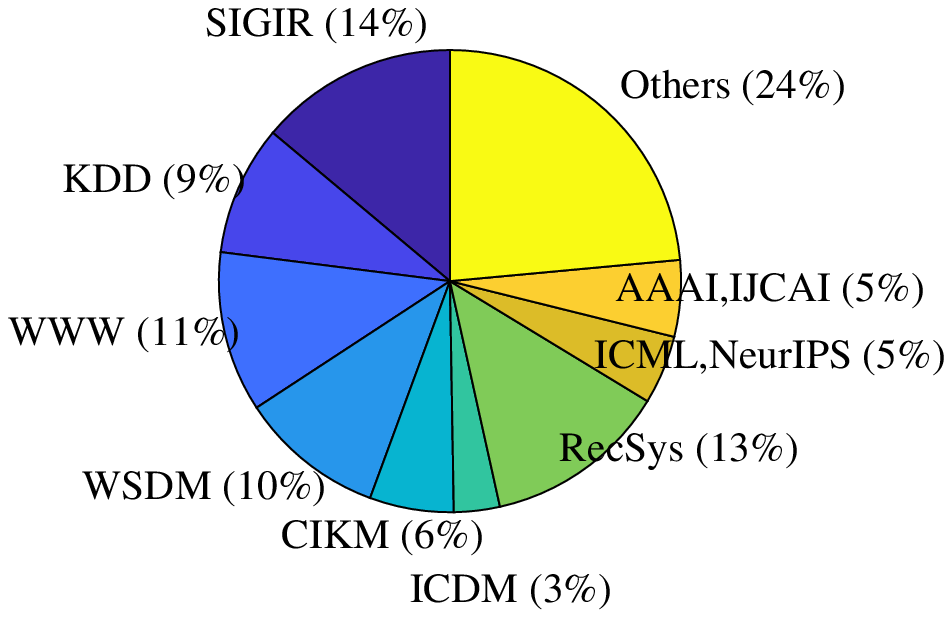}
}
\caption{The statistics of publications related to biases in RS with the publication year and venue. }\label{fg:pypb}
\end{figure}

\textbf{Increasing Importance of Biases in RS Research.} Recent years have seen a surge of research effort on  recommendation biases. Figure \ref{fg:pypb} shows the number of related papers in top venues increases significantly since the year of 2015. The prestigious international conference on information retrieval, SIGIR, has organized specific sessions in 2020 and 2021 to discuss topics on bias elimination\footnote{\url{http://www.sigir.org/sigir2020/schedule/};   \url{https://sigir.org/sigir2021/schedule/}}. SIGIR even presents the Best Paper award to the paper on this topic in 2018~\cite{canamares2018should}, 2020~\cite{DBLP:conf/sigir/MorikSHJ20} and 2021 ~\cite{DBLP:conf/sigir/Oosterhuis21,zhang2021causal}, respectively. The conferences Recsys and WWW also organized tutorial on this topic in 2021 \cite{chen2021bias,west2021summary}. Biases not only draw increasing attention from the information retrieval academia, but also from the industry. For example, one competing task of KDD Cup 2020 organized by Alibaba is to handle the long-tail bias in E-commerce recommendation\footnote{\url{https://tianchi.aliyun.com/competition/entrance/231785/introduction}}.

\textbf{Necessity of this Survey.} Although many papers are published on this topic recently, to the best of our knowledge, none of them has provided a global picture of the RS biases and corresponding debiasing techniques. Particularly, we find that current studies on this topic are rather fragmented --- despite the wide usage of the terminology ``bias'' in the literature, its definition is usually vague and even inconsistent across papers. For example, some work use ``selection bias'' to denote the bias of observed rating values~\cite{saito2020asymmetric}, while others use ``observational bias'' to refer to the same meaning instead~\cite{hernandez2014probabilistic}. More confusingly, the same terminology ``selection bias'' has been conceptualized differently in different publications \cite{wang2016learning, saito2020asymmetric, ovaisi2020correcting}. Moreover, a considerable number of researchers do not explicitly mention ``bias'' or ``debias'' in the paper (e.g. \cite{liang2016modeling,chen2018modeling,DBLP:conf/www/WangX000C20}), but they indeed address one type of biases in RS; these significant related work is difficult to be retrieved by the researchers interested in the bias topic.
Given the increasing attention of biases in RS, the rapid development of debiasing techniques, and the flourishing but fragmented publications,
we believe it is the right time to present a survey of this area, so as to benefit the successive researchers and practitioners to understand current progress and further work on this topic.

\textbf{Difference with Existing Surveys.}
A number of surveys in recommendation have been published recently, focusing on different perspectives of RS.
For example, \cite{zhang2018explainable} reviews explainable recommendation, \cite{tarus2018knowledge} reviews knowledge-based recommendation,
\cite{zhang2019deep} and \cite{zhao2019deep} summarize the recommendation methods based on deep learning and reinforcement learning, respectively.
However, to our knowledge, the perspective of bias has not been reviewed in existing RS surveys.
There are some surveys on the bias issues, but they are not on the recommendation task. For example, \cite{blodgett2020language} recently reviews the bias issues in natural language processing, \cite{vella1998estimating} reviews the sample selection bias on model estimation, \cite{zhang2020fairness}  summarizes fairness in learning-based sequential decision algorithms. There are some surveys on the bias and fairness of general machine learning and artificial intelligence systems~\cite{del2020review,mehrabi2019survey,ntoutsi2020bias,shen2021towards}.
Comparing to the bias issues in other tasks, bias in RS has its own characteristics, requiring a new inclusive review and summary.
To this end, we make the following contributions in this survey:
\begin{itemize}
\item Summarizing seven types of biases in RS and providing their definitions and characteristics. Wherein, we specifically provide causality-based explanations for data biases to help the readers to better understand their nature.
\item Conducting a comprehensive review and providing a taxonomy of existing methods on recommendation debiasing, as well as discussing their strengths and weaknesses.
\item Identifying open challenges and discussing future directions to inspire more research on this topic.
\end{itemize}

\textbf{Papers Collection.} We collect over 180 papers that analyze the bias issues in recommendation or propose new debiasing methods. We first search the related top-tier conferences and journals to find related work, inculding WWW, WSDM, SIGIR, KDD, RecSys, CIKM, TOIS, TKDE, etc., with the keywords ``recommend'', ``collaborative filtering'', ``ranking'' or ``search'' combined with ``bias'', ``fairness'' or ``exposure'' from the year 2010 to 2021.
We then traverse the citation graph of the identified papers, retaining the papers that focus bias in RS. Figure \ref{fg:pypb} illustrates the statistics of collected papers with the publication time and venue.

\textbf{Survey Audience and Organization.}
This survey is beneficial for the following researchers and practitioners in RS: 1) who are new to the bias issues and look for a handbook to fast step into this area, 2) who are confused by different bias definitions in the literature and need a systematic study to understand the biases, 3) who want to keep up with the state-of-the-art debiasing technologies in RS, and 4) who face bias issues in building recommender systems and look for suitable solutions.
The rest of the survey is organized as follows: Section 2 introduces the preliminaries of RS and the critical issue of feedback loop. Section 3 and 4 are the main content, which summarizes the seven types of biases and provides a taxonomy of debiasing technologies in RS. Section 5 discusses open challenges and future directions, and Section 6 concludes the survey.

\section{Preliminaries: Recommender System and Feedback Loop}
\subsection{Feedback Loop in Recommendation}

\begin{figure}[t!]
\centering
\includegraphics[width=0.95\textwidth]{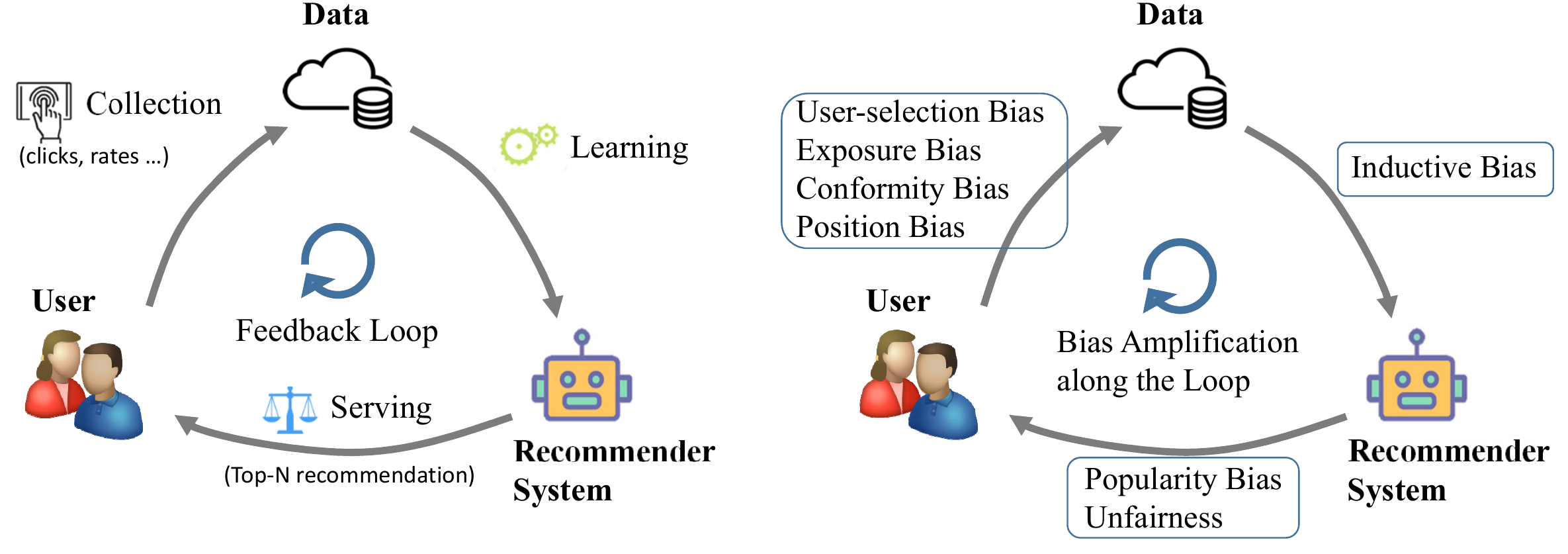}
 \caption{Feedback loop in recommendation, where biases occur in different stages.}
\label{fg:allbias}
\end{figure}

From a bird's-eye view, we can abstract the lifecycle of recommendation as a feedback loop among three key components: User, Data, and Model.
As Figure~\ref{fg:allbias} shows, the feedback loop consists of three stages:
\begin{itemize}
    \item \textbf{User$\rightarrow$Data} (Collection), which indicates the phase of collecting data from users, including user-item interactions and other side information (\eg user profile, item attributes, and contexts).
    \item \textbf{Data$\rightarrow$Model} (Learning), which represents the learning of recommendation models based on the collected data. At its core is to derive user preference from historical interactions, and predict how likely a user would adopt a target item. Extensive studies have been conducted over past decades.
    \item \textbf{Model$\rightarrow$User} (Serving), which returns the recommendation results to users, so as to satisfy the information need of users. This stage will affect the future behaviors and decisions of users.
\end{itemize}
Through this loop, users and the RS are in a process of mutual dynamic evolution, where personal interests and behaviors of users get updated via recommendation, and the RS can lead to a self-reinforcing pattern by leveraging the updated data.

\subsection{Recommendation Task Formulation}
We use uppercase character (\eg $U$) to denote a random variable; lowercase character (\eg $u$) to denote its specific value; and the character in calligraphic font (\eg $\mathcal U$) to represent the space of the corresponding random variable. The probability distribution of a random variable is notated with $p(.)$, while the expectation of a function of a random variable $X$ is notated with $\mathbb E_{X}[.]$.

Suppose we have a recommender system with a user set $\Set U$ and an item set $\Set I$. We denote the collected interactions $D_T$ between the user set~$\Set{U}$ and item set~$\Set{I}$, as a list of user-item-label triplets $(u,i,r)$, which are drawn from an unknown training distribution $p_T(U,I,R)$. Here $r \in \Set R$ denotes the feedback label given by a user to an item. It can be explicit (\eg rating values) that directly reflects user preference on the rated item, or be implicit (\eg purchase, click, view) indicating whether the user $u$ is willing to interact with the item $i$. For better description, we use the notation $r_{ui}$ with the subscript denoting the ground-truth label of the user-item pair $(u,i)$, while use the $r^o_{ui}$ denoting the observed label of $(u,i)$ in training data. Besides, we define a Bernoulli random variable $S\in \{0,1\}$ indicating whether the instance is observed in $D_T$ and use notation $s_{ui}$ for a certain user-item pair (\ie $s_{ui}=1$ \textit{iff.} $(u,i,r^o_{ui})\in D_T$). The task of a recommendation system can be stated as follows: learning a recommendation model from the available dataset $D_T$ so that it can capture user preference and make a high-quality recommendation in the serving stage. Formally, let $\delta (.,.)$ denote the error function measuring the distance between the prediction and the ground truth label \footnote{Without loss of generalization, here we just present the point-wise loss. In fact, other types of losses can be extended straightforwardly if we regard the item pair or list as an instance.}. The goal of recommendation is to learn a parametric model $f: \Set U\times \Set I \to \Set R$ from $D_T$ to minimize the following \textit{True Risk}:
\begin{equation}
    \label{risk_equ}
    \begin{split}
        L(f)=\mathbb E_{(u,i,r)\sim p_E}[\delta (f(u,i),r)] \\
    \end{split}
\end{equation}
where $P_E(U,I,R)$ denotes the ideal unbiased data distribution for model testing. This distribution can be factorized as the product of the user-item pair distribution $p_E(U,I)$ (often supposed as uniform) and the factual preference distribution for each user-item pair $p_E(R|U,I)$. As such, the \textit{True Risk} is often written as a metric over all user-item pairs:
\begin{equation}
    \begin{split}
     L(f) = \frac{1}{{|\Set U||\Set I|}}\sum\limits_{u\in \Set U,i\in \Set I} \mathbb E_{{r_{ui}}\sim {p_E}(R|U,I)}[{\delta (f(u,i),{r_{ui}})}]
    \end{split}
\end{equation}

Since the ideal test distribution is not accessible, the learning is conducted on the training set $D_T$ by optimizing the following \textit{empirical risk}:
\begin{equation}
    \label{em_risk_equ}
    \begin{split}
        \hat L_T(f)=\frac{1}{|D_T|}\sum\limits_{(u,i):{s_{ui}} = 1} {\delta (f(u,i),r_{ui}^o)}  \\
    \end{split}
\end{equation}
If the training data and test data are identically and independently distributed (\aka \textit{i.i.d.} assumptions), the empirical risk $\hat L_T(f)$ would be an unbiased estimator of the true risk $L(f)$, \ie $\mathbb E_{p_T}[L_T(f)]=L(f)$. The PAC learning theory~\cite{haussler1990probably} states that the learned model will be approximately optimal if we have sufficiently large training data set.

\section{Bias in Recommendation}
In this section, we first summarize and position different types of biases in the feedback loop, as illustrated in Figure~\ref{fg:allbias} and Table \ref{tb:ch}.
We then present in-depth analyses of their relations and discuss how they are intensified along the feedback loop.


\begin{table*}[t!]
\centering
\scriptsize
\caption{The characteristics of seven types of biases in recommendation and the bias amplification in loop.}
\label{tb:ch}
\begin{tabular}{|l|l|l|l|l|}
\hline
Types           & Stages in Loop               & Cause                                                                                & Effect                                                                                                       & Major solutions                                                                                                   \\ \hline\hline
Selection Bias  & User$\rightarrow$Data      & Users' self-selection                                                                & \begin{tabular}[c]{@{}l@{}}Skewed observed \\ rating distribution\end{tabular}                               & \begin{tabular}[c]{@{}l@{}}Data Imputation; Propensity Score; \\ Joint Generative Model; Doubly Robust Model\end{tabular}                \\ \hline
Exposure Bias   & User$\rightarrow$Data      & \begin{tabular}[c]{@{}l@{}}Item Popularity;\\ Intervened by systems; \\ User behavior and background  \\ \end{tabular}    & \begin{tabular}[c]{@{}l@{}}Unobserved interactions \\  do not mean negative\end{tabular}                                      & \begin{tabular}[c]{@{}l@{}}Giving confidence weights by heuristic, \\ sampling or exposure-based model; \\ Propensity Score; Causality-based Model\end{tabular} \\ \hline
Conformity Bias & User$\rightarrow$Data      & Conformity                                                                           & Skewed interaction labels                                                                                       & Modeling social or popularity effect                                                                              \\ \hline
Position Bias   & User$\rightarrow$Data      & \begin{tabular}[c]{@{}l@{}}Trust top of lists;\\ Exposed to top of lists\end{tabular} & \begin{tabular}[c]{@{}l@{}}Unreliable positive\\  data\end{tabular}                                          & Click models; Propensity Score; Trust-aware Model                                                                                    \\ \hline
Inductive Bias  & Data$\rightarrow$Model              & Added by researchers or engineers                                                                & \begin{tabular}[c]{@{}l@{}}Better generalization, \\ lower variance or \\ Faster recommendation\end{tabular} & -                                                                                                                 \\ \hline
Popularity Bias & Model$\rightarrow$User              & \begin{tabular}[c]{@{}l@{}}Algorithm and unbalanced data \end{tabular}                                                                      & Matthew effect                                                                                               & \begin{tabular}[c]{@{}l@{}}Regularization; Adversarial Learning;\\ Causal Graph \end{tabular}                      \\ \hline
Unfairness      & Model$\rightarrow$User              & \begin{tabular}[c]{@{}l@{}}Algorithm and unbalanced data \end{tabular}                                                                       & Unfairness for some groups                                                                                   & \begin{tabular}[c]{@{}l@{}}Rebalancing; Regularization; \\ Adversarial Learning; Causal Modeling\end{tabular}     \\ \hline \hline
\begin{tabular}[c]{@{}l@{}}Bias amplification \\ in Loop\end{tabular}    & All                                       & Feedback loop                                                                        & Enhance and spread bias                                                                                      & \begin{tabular}[c]{@{}l@{}}Break the loop by collecting random \\ data or using reinforcement learning\end{tabular}                   \\ \hline
\end{tabular}
\end{table*}

\subsection{Bias in Data}
As the data, of user interactions, are observational rather than experimental, biases are easily introduced into the data.
They typically stem from different subgroups of data, and make the recommendation models capture these biases and even scale them, thereby leading to systemic racism and suboptimal decisions. In this subsection, we would first give the general definition of the data bias and then categorize it into four groups: selection bias, conformity bias, exposure bias and position bias.


\subsubsection{Definition of data bias} The \textit{i.i.d.} assumption lays a foundation for recent learning-based methods to generalize well on the test environment. However, this assumption may not hold in real recommender systems. Typically, the data collection process in RS is observational rather than experimental. The sample selection or user decision would inevitably be affected by many undesirable factors, such as the exposure mechanism of RS or public opinions, making the training data distribution deviate from test distribution. Training data only gives a skewed snapshot of user preference, making the recommendation model sink into sub-optimal result. We name such notorious distribution deviation phenomenon as data bias:


\begin{itemize}
    \item\textbf{Data Bias.} \emph{The distribution for which the training data is collected is different from the ideal test data distribution.}
\end{itemize}

\begin{figure}[t!]
\centering
\includegraphics[width=0.95\textwidth]{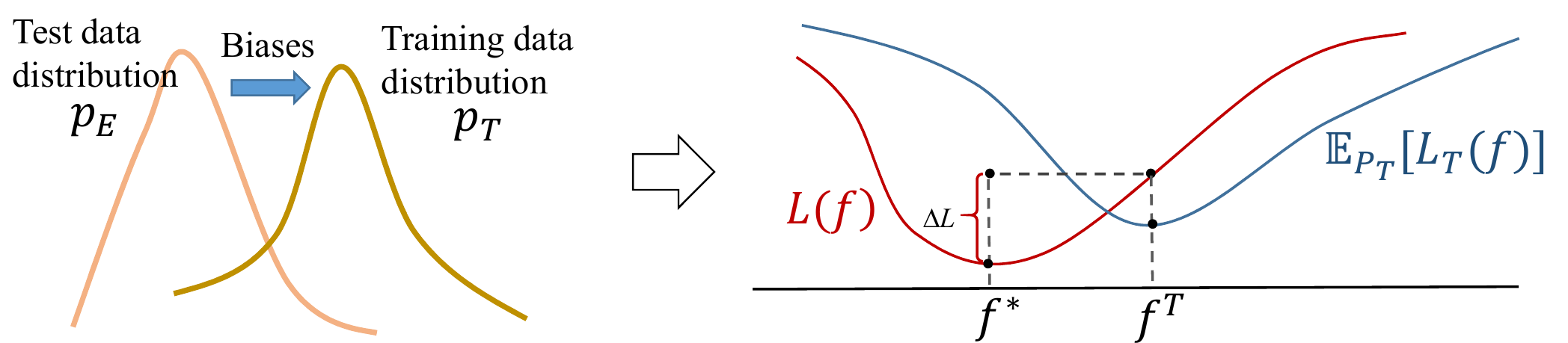}
 \caption{illustration of the data bias and its negative effect on model training.}
\label{fg:gap}
\end{figure}

Figure \ref{fg:gap} illustrates the data bias and its negative effect. Bias distorts training distribution, causing the model flow towards wrong direction. The red curve denotes the true risk function for testing, while the blue curve denotes the expected empirical risk function for training. As the two risks are expected over different distribution, they will behave rather differently even in their optimum (\ie $f^*$ versus $f^T$). It means that even if a sufficiently large training set is provided and the model arrives at empirical optimal point $f^T$, there still exists a certain gap $\Delta L$ between the optimum $L(f^*)$ and the empirical one $L(f^T)$. Blindly fitting a recommendation model without considering the inherent data bias will result in inferior performance.

In the following, we will introduce four types of data biases, with providing their definitions and characteristics. We also provide causality-based explanations for each bias to help the readers to better understand its nature.

\begin{figure}[t!]
\centering
 \subfigure[Randomly-selected items]{
\includegraphics[width=0.3\textwidth]{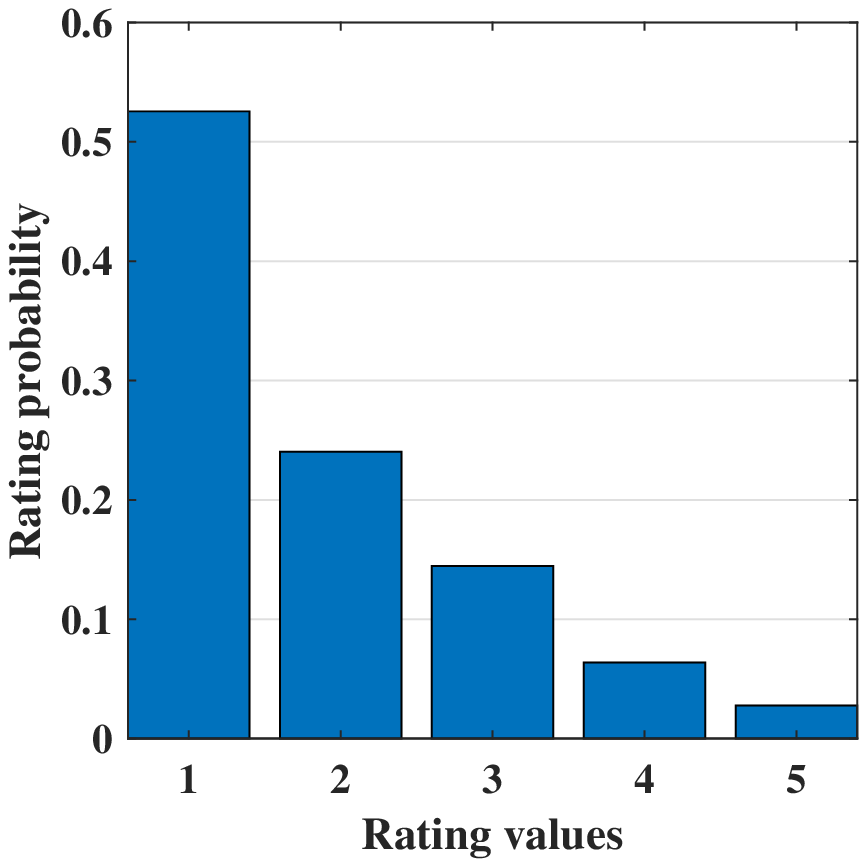}
}
\subfigure[User-selected items]{
\includegraphics[width=0.3\textwidth]{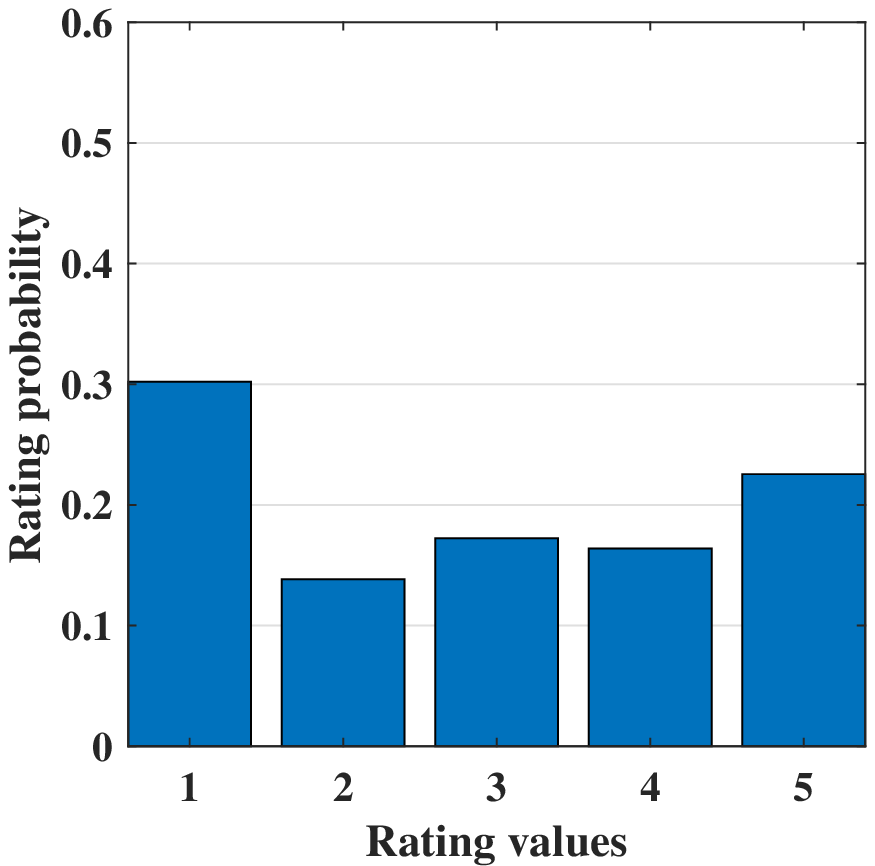}
}
 \caption{Distribution of rating values for randomly selected items and user-selected items. The data is from \cite{marlin2007collaborative} with permission.}
\label{mnar}
\end{figure}

\subsubsection{Selection Bias}
Selection bias originates from users' numerical ratings on items (\ie explicit feedback), which is defined as:
\begin{itemize}
    \item\textbf{Selection Bias.} \emph{Selection Bias happens as users are free to choose which items to rate, so that the observed ratings are not a representative sample of all ratings. In other words, the rating data is often missing not at random (MNAR).}
\end{itemize}

\begin{figure}[t!]
\centering
\includegraphics[width=0.95\textwidth]{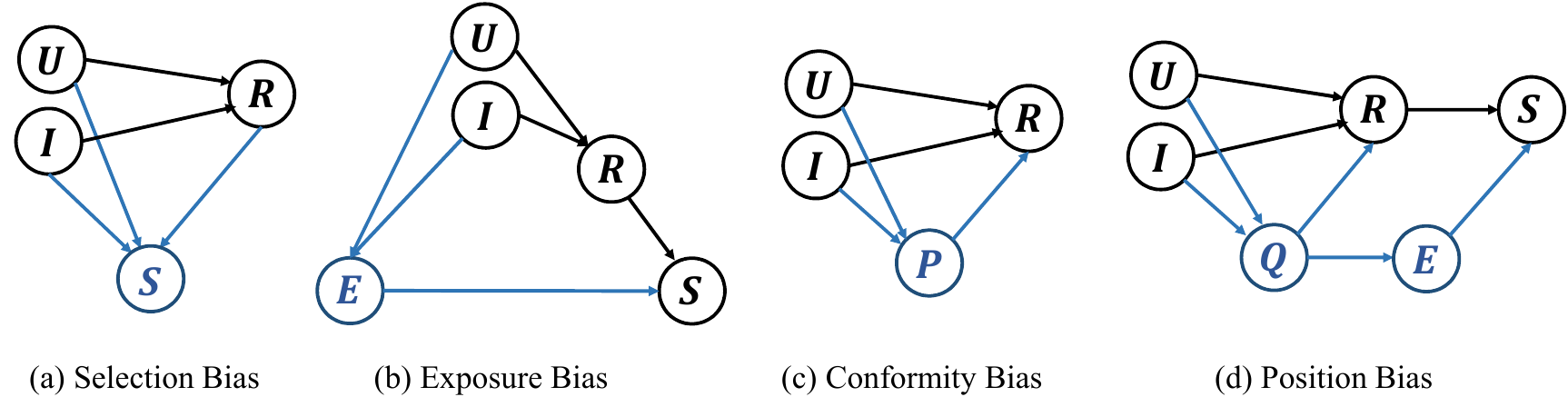}
 \caption{Causal graphs of four types of data bias. $U$: user; $I$: item; $R$: feedback label; $S$: observation variable (S=1, observed; S=0, unobserved); $E$: exposure; $P$: public opinion; $Q$: item position.}
\label{fg:cabias}
\end{figure}

Prior study conducted by Marlin~\etal~ \cite{marlin2007collaborative} offers compelling evidence to show the existence of selection bias in the rating data.
In particular, they conducted a user survey to gather the user ratings to some randomly-selected items, as a comparison with that to conventional user-selected items.
Figure \ref{mnar} summarizes the comparison and offers two findings:
1) users tend to select and rate the items that they like;
and 2) users are more likely to rate particularly bad or good items.
These results suggest that selection bias is inherent in the observed data, since they are missing not at random. The distribution of observed rating data is different from the distribution of all ratings~\cite{hernandez2014probabilistic,steck2013evaluation}.

\textbf{Causality-based Explanation.} Figure \ref{fg:cabias}(a) illustrates the generative process of the observed rating data. The links $(U,V) \to R$ represent the causal effect of the features of the user $U$ and item $I$ on their interaction label $R$, for which the recommendation model aims to estimate. In observational study, the collected rating data is not evenly presented and the variables $U,I,R$ would affected the observation of the instance ($S=1$, observed; $S=0$, unobserved). This mechanism makes the distribution of the observed rating data (\ie ${p_T} \equiv p(U,I,R|S=1)$) inconsistent with the ideal test distribution (\ie ${p_E} \equiv p(U,I,R)$). The model that directly learned on observed data would suffer.

From another perspective, the causal graph shows two sources of association between the causes $(U,I)$ and the outcome $R$: 1) the desirable causal effect $(U,V) \to R$; 2) the collision path \cite{pearl2018book} $(U,I) \to S \leftarrow R$ that links $(U,I)$ and $R$ through their common (conditioned on) effects $S=1$. The analyses conditioned on $S=1$ would create spurious association between $(U,I)$ and $R$. The model learned on observed data would capture skewed patterns.

\subsubsection{Exposure Bias}

Implicit feedback is widely used in recommendation, which reflects natural behaviors of users, such as purchases, views, clicks. Distinct from explicit feedback that offers numerical ratings, implicit feedback only provides partial signal of positive. As the knowledge about what the user dislikes is not available, the learning must rely on unobserved interactions, mining the negative signal from them. Exposure bias happens in such one-class data, which is defined as:

\begin{itemize}
\item \textbf{Exposure Bias. } \textit{Exposure bias happens as users are only exposed to a part of specific items so that unobserved interactions do not always represent negative preference.}
\end{itemize}
In particular, an unobserved interaction between a user and an item can be attributed to two possible reasons: 1) the item does not match user interest; and 2) the user is unaware of the item. Hence, ambiguity arises in the interpretation of unobserved interactions. The inability to distinguish real negative interactions (e.g. exposed but uninterested) from the potentially-positive ones (e.g. unexposed) will result in severe biases. Previous studies have investigated several dimensions of data exposure:
1) Exposure is affected by the policy of the previous recommender systems, which controls what items to show~\cite{liu2020general}.
Hence, some recent works~\cite{liu2020general} also name such ``exposure bias'' as ``previous model bias''.
2) As users may actively search and find the items of interest, the selection of users is a factor of exposure~\cite{ovaisi2020correcting,wang2016learning}, and makes highly relevant items more likely to be exposed.
Hence, in this scenario, ``exposure bias'' is named as ``user-selection bias''.
3) The background of users is another factor to expose items, such as social friends~\cite{chen2019samwalker}, communities that they belong to~\cite{chen2020fast}, and geo locations.
4) Popular items are more likely to be seen by users. Hence, such ``popularity bias'' is another form of ``exposure bias''\cite{DBLP:journals/corr/abs-2006-11011}.
In order to facilitate readers and prevent concept confusion, we use the unified standard definition, ``exposure bias'', throughout this paper, rather than the separated definitions of the aforementioned factors.

\textbf{Causality-based Explanation.} Figure \ref{fg:cabias}(b) illustrates the generative process of the collected implicit feedback data: the links $(U,I) \to R$ represent the causal effect of user/item features on the feedback, for which the recommender aims to estimate; the link $R\to S$ represents the missing mechanism in implicit feedback --- only positive interactions are observed $p(R=1|S=1)=1$. As the negative instances are not available, the learning must resort to the unobserved interactions. Recent work on implicit feedback would leverage the observation variable $S$ as a surrogate label --- \ie marking observed interactions as positive while unobserved as negative. It is rational in ideal scenario as the distribution $p(R|U,I)$ is equal to $P(S|U,I)$. However, due to the wide existence of the exposure bias in practical, the equation does not hold. It can be understood from the paths $(U,I)\to E \to S$, where the user/item features would affect whether the item is exposed to the user (marked as a variable $E$); and $E$ would further distort the distribution $P(S|U,I)$, as a user could only generate interactions on the exposed items \ie $P(S=0|E=0)=1$.

From another perspective, Figure \ref{fg:cabias}(b) shows two sources of associations between the variables $(U,I)$ and $S$. 1) the desirable causal effect along the $(U,V) \to R \to S$; 2) the spurious association created by the exposure bias through the paths $(U,V) \to E \to S$. The correlations captured by the recommendation model may fail to reflect the true preference.


\subsubsection{Conformity Bias} Different from the aforementioned biases that contribute on the data observation, conformity bias distorts user judgment, which is defined as follow:

\begin{itemize}
    \item \textbf{Conformity Bias. } \textit{Conformity bias happens as users tend to behave similarly to the others in a group, even if doing so goes against their own judgment, making the feedback do not always signify user true preference.}
\end{itemize}
For example, influenced by high ratings of public comments on an item, one user is highly likely to change her low rate, avoiding being too harsh~\cite{wang2014amazon,liu2016you}. Such phenomenon of conformity is common and cause biases in user ratings \cite{lederrey2018sheep}.
As shown in Krishnan \etal~ \cite{krishnan2014methodology}, user ratings follow different distributions when users rate items before or after being exposed to the public opinions.
Moreover, conformity bias might be caused by social influence, where users tend to behave similarly with their friends\cite{ma2009learning,tang2012mtrust,chaney2015probabilistic,wang2017learning}.
Hence, the observed interactions are skewed and might not reflect users' real preference on items \cite{liu2016you}.

\textbf{Causality-based Explanation.}
As illustrated in Figure \ref{fg:cabias}(c), conformity bias can be understood from the additional causal path $I\to P \to R$.  Public opinion $P$ (\eg popularity, averaged rating) depends on item property ($I$) and impacts user judgment $R$. This undesirable phenomenon would distort the conditional distribution $p(R|U,I)$, making the training distribution deviate from reflecting user true preference.

From another perspective, conformity bias creates spurious association between $(U,I)$ and $R$ through the path $I\to P \to R$. The model trained on observed data would easily capture spurious association, leading to poor performance.

\subsubsection{Position Bias} Position bias is very common in recommendation, particularly in advertisement system or search engine:
\begin{itemize}
\item \textbf{Position Bias. } \textit{Position bias happens as users tend to interact with items in higher position of the recommendation list regardless of the items' actual relevance so that the interacted items might not be highly relevant.}
\end{itemize}
Here ``relevance'' is widely used in the field of information retrieval, which denotes how the items are preferred by the users. Popularity bias happens in implicit feedback data and describes a tendency of users to notice or interact with items in certain positions of lists with higher probability, regardless of the items' actual relevance\cite{collins2018study}.
For example, recent studies on eye tracking demonstrate that users are less likely to browse items that are ranked lower in vertical lists, while they only examine the first few items at the top of lists~\cite{joachims2007evaluating,joachims2017accurately}.
Moreover, Maeve \etal~ \cite{o2006modeling} shows that users often trust the first few results in the lists and then stop assessing the rest, without evaluating the entire list holistically~\cite{klockner2004depth}.
As such, the data collected from user feedback towards the recommended lists may fail to reflect user preference faithfully~\cite{collins2018study}.

\textbf{Causality-based Explanation.} Figure \ref{fg:cabias}(d) illustrates the generative process of users' feedback on recommendation lists, where each item is companied with a position $Q$ displayed in the previous recommendation list. The effect of position $Q$ on outcome $S$ is multi-folded: 1) along the path $Q\to E \to S$, the display position would impact the probability that the item is exposed to the user; 2) along the path $Q\to R \to S$, the position would also hinder users' own judgement, as users trust the recommender system and may over-estimate the relevance of the highly-ranked items. As such, position bias is quite complex and would skew both the data observation $p(S|U,I,R)$ as well as the user judgment $p(R|U,I)$, making the training distribution deviate significantly from the ideal test one.

From another perspective, position bias creates two spurious associations between $(U,I)$ and $S$ through the paths $(U,I)\to Q \to E \to S$ and $(U,I)\to Q \to R \to S$, which should be conquered.

\subsection{Bias in Model}
Bias is not always harmful. In fact, a number of inductive biases have been added deliberately into the model design to achieve some desirable characteristics:
\begin{itemize}
\item \textbf{Inductive Bias. } \textit{Inductive bias denotes the assumptions made by the model to better learn the target function and to generalize beyond training data.}
\end{itemize}
The ability to generalize the prediction to unseen examples is the core of machine learning. Without assumptions on the data or model, generalization cannot be achieved since the unseen examples may have an arbitrary output space.
Similarly, building a RS needs to add some assumptions on the nature of the target function.  For example, Johnson ~\etal \cite{johnson2014logistic} assumes an interaction can be estimated by embedding inner product, while He \etal~\cite{he2017neural} adopts the neural network as its better generalization. Besides target function, inductive bias have been added in other aspects. An example is the adaptive negative sampler\cite{rendle2014improving,wang2017irgan,park2019adversarial,ding2019reinforced}, which aims to over-sample the ``difficult'' instances in order to increase learning speed, even though the resultant loss function will differ significantly from the original. Another example is the discrete ranking model\cite{zhang2016discrete,lian2017discrete,zhou2012learning} which embeds user and items as binary codes to improve the efficiency of recommendation, which is at the expense of sacrificing the representation ability.

\subsection{Bias and Unfairness in Results}

Besides the aforementioned biases in data or model, two important biases in recommendation results have been studied, which are defined as follows:


\begin{itemize}
\item \textbf{Popularity Bias. } \textit{Popular items are recommended even more frequently than their popularity would warrant~\cite{abdollahpouri2020multi}.}
\end{itemize}

\begin{figure}[t!]
  \centering
  \includegraphics[width=0.5\textwidth]{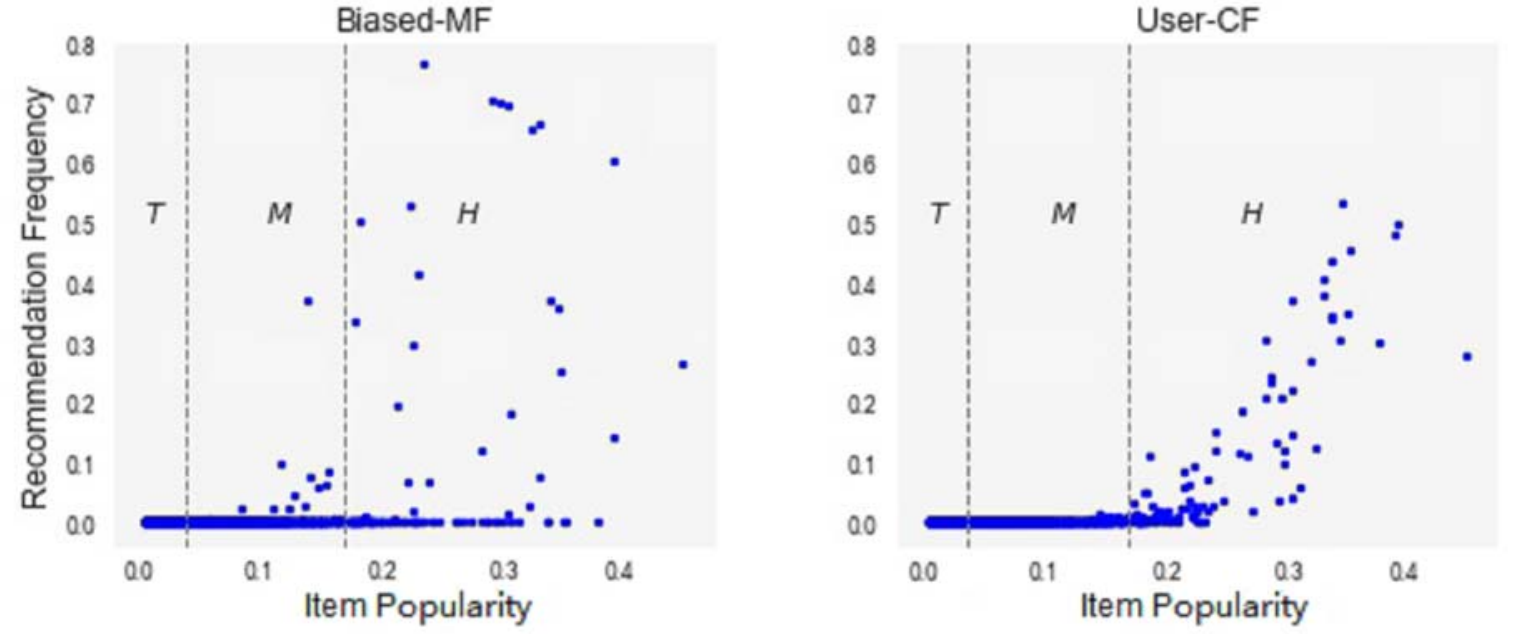}
  \caption{Item popularity VS. recommendation frequency  (Biased-MF~\cite{koren2009matrix} and User-CF~\cite{aggarwal2016neighborhood}), where items were classified into three different groups: H denoting the set of most popular items that take up around 20\% entire ratings, T denoting the set of most unpopular items that take up 20\% entire ratings and M denotes the rest. The figure was reproduced from ~\cite{abdollahpouri2020multi} with authors' permission.}
  \label{pop_al1}
\end{figure}

The long-tail phenomenon is common in RS data: in most cases, a small fraction of popular items account for the most of user interactions \cite{abdollahpouri2020multi}. When trained on such long-tailed data, the  model usually gives higher scores to popular items than their ideal values while simply predicts unpopular items as negative. As a result, popular items are recommended even more frequently than their original popularity exhibited in the dataset. Popularity bias has been empirically verified by Abdollahpouri \etal ~\cite{abdollahpouri2020multi,abdollahpouri2019unfairness}. Figure~\ref{pop_al1} shows relationship between item popularity and recommendation frequency. We can find most of recommended items are located at high popularity area (H). In fact, they are recommended to a much greater degree than even what their initial popularity warrants~\cite{abdollahpouri2020multi}.

Ignoring the popularity bias results in many issues:
1) It decreases the level of personalization and hurts the serendipity.
Since the preferences of different users are diverse, always recommending popular items will hurt user experience, especially for the users favoring niche items.
2) It decreases the fairness of the recommendation results \cite{abdollahpouri2020connection}.
Popular items are not always of high quality. Over-recommending popular items will reduce the visibility of other items even if they are good matches, which is unfair.
3) Popular bias will further increase the exposure opportunities of popular items, making popular items even more popular -- the collected data for future training becomes more unbalanced, raising the so-called ``Matthew effect'' issue \cite{DBLP:conf/kdd/Zhu0ZC21}.

Another type of bias arises in the recommendation results is unfairness. Fairness has attracted increasing attention in recent years. A consensual definition of fairness is ``\emph{absence of any prejudice or favoritism towards an individual or a group based on their intrinsic or acquired traits}''~\cite{DBLP:journals/corr/abs-1908-09635}, and the unfairness can be defined as follow:

\begin{itemize}
\item \textbf{Unfairness.} \textit{The system systematically and unfairly discriminates against certain individuals or groups of individuals in favor others~\cite{DBLP:journals/tois/FriedmanN96}.}
\end{itemize}

Unfairness issue has been an obstacle to making recommender systems more entrenched within our society.
In particular, based on attributes like race, gender, age, education level, or wealth, different user groups are usually unequally represented in data.
When training on such unbalanced data, the models are highly likely to learn these over-represented groups, reinforce them in the ranked results, and potentially result in systematic discrimination and reduced the visibility for disadvantaged groups (\eg under-representing the minorities, racial or gender stereotypes) \cite{DBLP:conf/recsys/LinSMB19}.
For example, in the context of job recommendation, previous work~\cite{lambrecht2019algorithmic,datta2015automated} found that, compared to men, women saw less ads about high paying jobs and career coaching services, which is caused by gender imbalance.
Analogously, friend recommendation in social graphs may reinforce historical biases towards a majority and prevent minorities from being social influencers with high reach~\cite{DBLP:conf/www/StoicaRC18,karimi2018homophily}.
Another similar issue has been found in book recommendation, where the methods prefer recommending books of male authors~\cite{DBLP:conf/recsys/EkstrandTKMK18}.
Analyzing the unfairness issue inherent in recommendation is therefore becoming essential and desirable.

%
%
%

\begin{figure}[t!]
  \centering
  \includegraphics[width=0.7\textwidth]{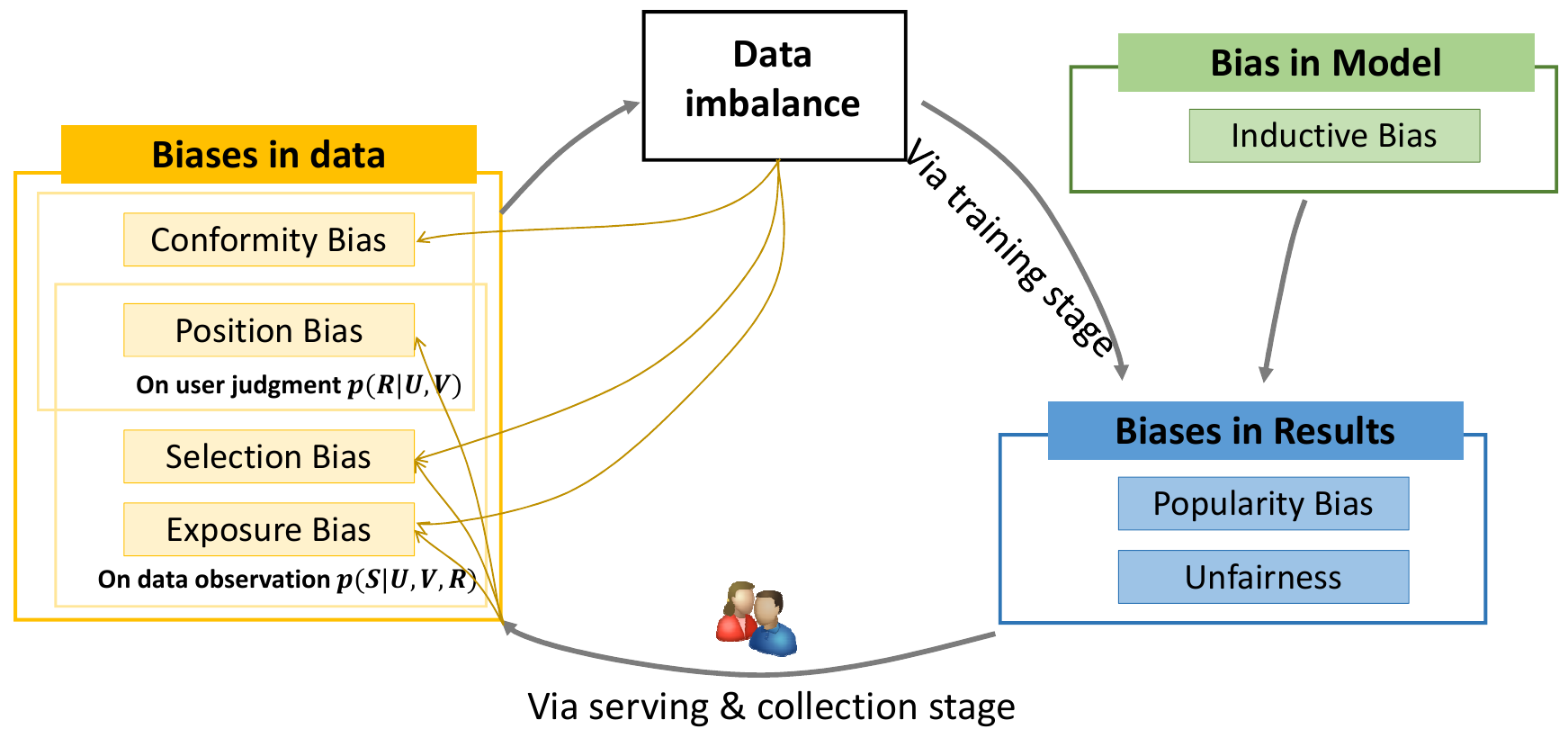}
  \caption{Relations between seven types of biases.}
  \label{fg:relations}
\end{figure}

\subsection{Feedback Loop Amplifies Biases}

Real-world recommender systems usually create a pernicious feedback loop. Previous subsections summarize the biases occurred in different stages of the loop, while these biases could be further intensified over time along the loop. To better understand this effect, figure \ref{fg:relations} illustrates the relations of seven types of biases. Data biases would incur or intensify the data imbalance, exacerbating bias issues in recommendation results (\textit{Data bias $\to$ Data imbalance $\to$ Bias in Results}); while the biased recommendations would in turn impact the decisions, exposure, and selections of users, reinforcing the biases in users' future behaviors (\textit{Bias in Results $\to$ Data Bias}). Taking the position bias as an example, top items typically benefit from a greater volume of traffic, which in turn increases their ranking prominence and the volume of traffic they receive, resulting in a rich-get-richer scenario\cite{o2006modeling}. Many researchers also study the impact of feedback loop on the popularity bias \cite{mansoury2020feedback,chaney2018algorithmic,jannach2015recommenders}. Their simulated results show that feedback loop will amplify popularity bias, where popular items become even more popular and non-popular items become even less popular. These amplified biases also will decrease the diversity and intensify the homogenization of users, raising the so-called ``echo chambers'' or ``filter bubbles'' \cite{jiang2019degenerate,ge2020understanding}.

Interestingly, figure \ref{fg:relations} also shows that some data biases can be self-reinforced (\textit{Data bias $\to$ Data imbalance $\to$Data bias}). Taking conformity bias as an example, as users tend to behave similarly to the major groups, the extent of the data imbalance would increase with time going by, which in turn impacts and exacerbates the conformity bias. Similar situations apply to the selection bias and exposure bias which are also affected by the data imbalance.

\begin{table*}[!htbp]
\tiny
\caption{A lookup table for the reviewed methods for recommendation debiasing.}
\begin{tabular}{|l|ll|l|l|l|l|}
\hline
Addressed issues                                                                          & \multicolumn{2}{l|}{Categories}                                           & How?                                                                                                                 & Strengths?                                                                               & Weaknesses?                                                                                                      & Publications \\ \hline
\multirow{6}{*}{Selection Bias}                                                           & \multicolumn{1}{l|}{\multirow{2}{*}{Evaluator}} & Propensity Score        & Weight the data                                                                                                      & \begin{tabular}[c]{@{}l@{}}General,\\ Theoretical-soundness\end{tabular}                 & \begin{tabular}[c]{@{}l@{}}Requiring proper \\ propensities\end{tabular}                                         & \cite{schnabel2016recommendations}             \\ \cline{3-7}
                                                                                          & \multicolumn{1}{l|}{}                           & ATOP                    & Specific design                                                                                                      & Theoretical-soundness                                                                    & \begin{tabular}[c]{@{}l@{}}Requiring two \\ strong assumptions\end{tabular}                                      &  \cite{steck2010training}            \\ \cline{2-7}
                                                                                          & \multicolumn{1}{l|}{\multirow{4}{*}{Training}}  & Joint Generative Model  & Model missing mechanism                                                                                               & Explainable                                                                              & \begin{tabular}[c]{@{}l@{}}Requiring assumptions \\ on data generation,\\ Hard to train\end{tabular}             &  \begin{tabular}[c]{@{}l@{}} \cite{hernandez2014probabilistic,marlin2007collaborative,marlin2009collaborative,jiawei2018social,wang2018deconfounded,kim2014bayesian}\\ \cite{yang2015boosting}  \end{tabular}          \\ \cline{3-7}
                                                                                          & \multicolumn{1}{l|}{}                           & Data Imputation         & Impute pseudo-labels                                                                                                 & Simple                                                                                   & \begin{tabular}[c]{@{}l@{}}Highly sensitive \\ to pesduo-labels\end{tabular}                                     &  \cite{steck2013evaluation,steck2010training,saito2020asymmetric}            \\ \cline{3-7}
                                                                                          & \multicolumn{1}{l|}{}                           & Propensity Score        & Weight the data                                                                                                      & Theoretical-soundness                                                                    & \begin{tabular}[c]{@{}l@{}}High variance,\\ Requiring proper \\ propensities\end{tabular}                        &   \cite{schnabel2016recommendations,wang2021combating}           \\ \cline{3-7}
                                                                                          & \multicolumn{1}{l|}{}                           & Doubly Robust Model     & Impute+Weight                                                                                                        & \begin{tabular}[c]{@{}l@{}}Theoretical-soundness,\\ Robust\end{tabular}                  & \begin{tabular}[c]{@{}l@{}}Requiring proper \\ propensities\\ or pesduo-labels\end{tabular}                      &   \cite{wang2019doubly}           \\ \hline
\multirow{2}{*}{Conformity Bias}                                                          & \multicolumn{2}{l|}{Modeling popularity influence}                        & \begin{tabular}[c]{@{}l@{}}Disentangle conformity\\ effect from user preference\end{tabular}                         & Explainable                                                                              & \begin{tabular}[c]{@{}l@{}}Requiring assumptions\\ on data generation\end{tabular}                               &    \cite{liu2016you,DBLP:journals/corr/abs-2006-11011,zhao2021popularity}          \\ \cline{2-7}
                                                                                          & \multicolumn{2}{l|}{Modeling social influence}                            & \begin{tabular}[c]{@{}l@{}}Disentangle social effect \\ from user preference\end{tabular}                            & Explainable                                                                              & \begin{tabular}[c]{@{}l@{}}Requiring assumptions \\ on data generation\end{tabular}                              &   \cite{ma2009learning,tang2012mtrust,chaney2015probabilistic,wang2017learning}           \\ \hline
\multirow{7}{*}{Exposure Bias}                                                            & \multicolumn{1}{l|}{Evaluator}                  & Propensity Score        & Weight the data                                                                                                      & Theoretical-soundness                                                                    & \begin{tabular}[c]{@{}l@{}}Requiring proper \\ propensities\end{tabular}                                         &   \cite{yang2018unbiased}            \\ \cline{2-7}
                                                                                          & \multicolumn{1}{l|}{\multirow{6}{*}{Training}}  & Heuristic               & \begin{tabular}[c]{@{}l@{}}Down-weight unobserved \\ data heurstically\end{tabular}                                  & Simple                                                                                   & \begin{tabular}[c]{@{}l@{}}Coarse-grained, \\ Heuristical\end{tabular}                                           &   \begin{tabular}[c]{@{}l@{}} \cite{hu2008collaborative,devooght2015dynamic,pan2009mind,Pan2008,he2016fast,yu2017selection} \\ \cite{li2010improving,saito2020unbiased} \end{tabular}           \\ \cline{3-7}
                                                                                          & \multicolumn{1}{l|}{}                           & Sampling                & \begin{tabular}[c]{@{}l@{}}Down-weight unobserved \\ data via sampling\end{tabular}                                  & Efficient                                                                                & \begin{tabular}[c]{@{}l@{}}Coarse-grained,\\ Requiring heuristic\\ or side information\end{tabular}              &    \cite{yu2017selection,rendle2009bpr,ding2018improved,ding2019reinforced,chen2019samwalker,wang2020reinforced}           \\ \cline{3-7}
                                                                                          & \multicolumn{1}{l|}{}                           & Exposure-based model    & \begin{tabular}[c]{@{}l@{}}Weight the data via \\ exposure model\end{tabular}                                        & \begin{tabular}[c]{@{}l@{}}Explainable,\\ Learn flexible weights\end{tabular}            & Hard to train                                                                                                    &   \cite{DBLP:conf/aaai/WangZYZ18, chen2018modeling,liang2016modeling,chen2019samwalker,chen2020fast}             \\ \cline{3-7}
                                                                                          & \multicolumn{1}{l|}{}                           & Propensity Scores       & Weight the observed data                                                                                             & Theoretical-soundness                                                                    & \begin{tabular}[c]{@{}l@{}}Requiring proper\\ propensities,\\ High variance\end{tabular}                         &    \cite{saito2020unbiased,zhu2020unbiased}          \\ \cline{3-7}
                                                                                          & \multicolumn{1}{l|}{}                           & Causality-based Methods & \begin{tabular}[c]{@{}l@{}}Remove spurious associations\\ via causal inference\end{tabular}                          & Explainable                                                                              & \begin{tabular}[c]{@{}l@{}}Requiring assumptions\\ on data generation\end{tabular}                               &    \cite{zhang2021causal,xu2021deconfounded,yang2021top,liu2021mitigating}          \\ \cline{3-7}
                                                                                          & \multicolumn{1}{l|}{}                           & Others                  & -                                                                                                                    & -                                                                                        & -                                                                                                                &   \begin{tabular}[c]{@{}l@{}} \cite{wang2016learning,ovaisi2020correcting,DBLP:conf/sigir/OvaisiVZ21,zhang2020large,ma2018entire} \\ \cite{wen2020entire,bao2020gmcm,damak2021debiased,schnabel2020debiasing}  \end{tabular}          \\ \hline
\multirow{3}{*}{Position Bias}                                                            & \multicolumn{2}{l|}{Click Models}                                         & \begin{tabular}[c]{@{}l@{}}Model the generative \\ process of clicks\end{tabular}                                    & Explainable                                                                              & \begin{tabular}[c]{@{}l@{}}Requiring assumptions\\ on data generation,\\ hard to train\end{tabular}              &   \begin{tabular}[c]{@{}l@{}} \cite{lin2021graph,craswell2008experimental,dupret2008user,chapelle2009dynamic,zhang2007comparing,guo2009click}\\ \cite{zhu2010novel,jin2020deep}      \end{tabular}     \\ \cline{2-7}
                                                                                          & \multicolumn{2}{l|}{Propensity Score}                                     & Weight the data                                                                                                      & Theoretical-soundness                                                                    & \begin{tabular}[c]{@{}l@{}}High variance,\\ Requiring proper \\ propensities,\\ Fail to model trust\end{tabular} &   \begin{tabular}[c]{@{}l@{}} \cite{agarwal2019general,joachims2017unbiased,schuth2016multileave,wang2016learning,raman2013learning,swaminathan2015batch,hofmann2013reusing}  \\ \cite{joachims2017unbiased,fang2019intervention,ai2018unbiased,wang2018position,vardasbi2020cascade,joachims2017unbiased,qin2020attribute} \\ \cite{chen2021adapting,guo2020debiasing,agarwal2019estimating}\end{tabular}          \\ \cline{2-7}
                                                                                          & \multicolumn{2}{l|}{Trust-aware Models}                                   & \begin{tabular}[c]{@{}l@{}}Introduce offset terms\\ to remove trust effect\end{tabular}                              & \begin{tabular}[c]{@{}l@{}}Theoretical-soundness,\\ Capture trust effect\end{tabular}    & \begin{tabular}[c]{@{}l@{}}High variance,\\ Requiring proper\\ propensities,\\ Fail to model trust\end{tabular}  &    \cite{agarwal2019addressing,vardasbi2020inverse}          \\ \hline
\begin{tabular}[c]{@{}l@{}}For multiple data biases\\ and their combinations\end{tabular} & \multicolumn{2}{l|}{Universal model}                                      & \begin{tabular}[c]{@{}l@{}}Transfer the knowledge \\ from unbiased data to \\ perform debiasing\end{tabular}         & \begin{tabular}[c]{@{}l@{}}Universal,\\ Adaptive\end{tabular}                            & \begin{tabular}[c]{@{}l@{}}Requiring a set of\\ unbiased data\end{tabular}                                       &  \cite{liu2020general,chen2021autodebias,bonner2018causal,lin2021transfer}            \\ \hline
\multirow{4}{*}{Popularity Bias}                                                          & \multicolumn{2}{l|}{Regularization}                                       & \begin{tabular}[c]{@{}l@{}}Introduce regularization\\ terms\end{tabular}                                             & Simple, Straightforward                                                                  & Possibly hurt accuracy                                                                                           &  \cite{zhu2021popularity,DBLP:conf/recsys/AbdollahpouriBM17,DBLP:conf/flairs/WasilewskiH16,DBLP:conf/recsys/KamishimaAAS14,DBLP:conf/sigir/ChenXLYSD20}             \\ \cline{2-7}
                                                                                          & \multicolumn{2}{l|}{Adversarial Learning}                                 & \begin{tabular}[c]{@{}l@{}}Leverage adversary to \\ bridge the gap between \\ niche and popular items\end{tabular}   & Balancing representation                                                                 & Possibly hurt accuracy                                                                                           &     \cite{DBLP:conf/cikm/KrishnanSSS18}         \\ \cline{2-7}
                                                                                          & \multicolumn{2}{l|}{Causal Graph}                                         & \begin{tabular}[c]{@{}l@{}}Leverage causal graph \\ to elucidate and mitigate \\ popularity bias\end{tabular}        & Explainable                                                                              & \begin{tabular}[c]{@{}l@{}}Requiring assumptions \\ on data generation\end{tabular}                              &   \cite{DBLP:journals/corr/abs-2006-11011,zhang2021causal,zhao2021popularity,wei2021model,DBLP:conf/kdd/WangF0WC21}           \\ \cline{2-7}
                                                                                          & \multicolumn{2}{l|}{Others}                                               & -                                                                                                                    & -                                                                                        & -                                                                                                                &    \cite{DBLP:conf/kdd/Bressan0PRT16,DBLP:conf/aies/Abdollahpouri19}           \\ \hline
\multirow{5}{*}{Unfairness}                                                               & \multicolumn{2}{l|}{Rebalancing}                                          & \begin{tabular}[c]{@{}l@{}}Directly balance the data\\ or recommendation results\end{tabular}                        & Straightforward                                                                          & Possibly hurt accuracy                                                                                           &    \begin{tabular}[c]{@{}l@{}} \cite{li2021user,DBLP:conf/kdd/PedreschiRT08,DBLP:conf/kdd/GeyikAK19,DBLP:conf/sigmod/AsudehJS019,DBLP:conf/sigir/BiegaGW18,DBLP:conf/cikm/ZehlikeB0HMB17,DBLP:conf/kdd/GeyikAK19} \\ \cite{DBLP:conf/kdd/SinghJ18,DBLP:journals/corr/abs-1809-09030,Fairwalk,DBLP:journals/corr/abs-2002-11442}   \end{tabular}           \\ \cline{2-7}
                                                                                          & \multicolumn{2}{l|}{Regularization}                                       & \begin{tabular}[c]{@{}l@{}}Formulate the fairness \\ criteria as a regularizer\end{tabular}                          & Straightforward                                                                          & Possibly hurt accuracy                                                                                           &    \begin{tabular}[c]{@{}l@{}} \cite{singh2019policy,yadav2021policy,LFR,DBLP:conf/recsys/KamishimaAAS12,kamishima2013efficiency,DBLP:conf/icdm/KamishimaAAS16,kamishima2017considerations} \\ \cite{DBLP:conf/nips/YaoH17,DBLP:journals/corr/YaoH17a,DBLP:conf/recsys/AbdollahpouriBM17,DBLP:conf/recsys/LinZZGLM17,burke2017balanced,DBLP:conf/kdd/BeutelCDQWWHZHC19}     \end{tabular}         \\ \cline{2-7}
                                                                                          & \multicolumn{2}{l|}{Adversarial Learning}                                 & \begin{tabular}[c]{@{}l@{}}Leverage adversary to \\ isolate the effect of\\ sensitive attributes\end{tabular}        & Fair representation                                                                      & Possibly hurt accuracy                                                                                           &   \cite{DBLP:conf/sigir/LiCXGZ21,wu2021learning,ALFR,CompositionalFairness,DBLP:conf/wsdm/BeigiMGAN020}              \\ \cline{2-7}
                                                                                          & \multicolumn{2}{l|}{Causal Modeling}                                      & \begin{tabular}[c]{@{}l@{}}Estimate fairness with\\ intervening sensitive \\ attributes\end{tabular}                 & \begin{tabular}[c]{@{}l@{}}Explainable, \\ Counterfactual fairness\end{tabular}          & \begin{tabular}[c]{@{}l@{}}Requiring assumptions\\ on data generation\end{tabular}                               &   \cite{DBLP:conf/aaai/ZhangB18,DBLP:conf/aaai/NabiS18,DBLP:conf/kdd/WuZW18,Counterfactual-Fairness,wu2019counterfactual,Counterfactual-Fairness}            \\ \cline{2-7}
                                                                                          & \multicolumn{2}{l|}{Others}                                               & -                                                                                                                    & -                                                                                        & -                                                                                                                &     \cite{li2021leave,islam2021debiasing,ge2021towards}         \\ \hline
\multirow{3}{*}{Loop effect}                                                              & \multicolumn{2}{l|}{Uniform Data}                                         & \begin{tabular}[c]{@{}l@{}}Intervene in the system \\ with a random logging policy\end{tabular}                      & \begin{tabular}[c]{@{}l@{}}Straightforward,\\ Effective\end{tabular}                     & \begin{tabular}[c]{@{}l@{}}Hurting the user \\ experience and the \\ system profit\end{tabular}                  &    \begin{tabular}[c]{@{}l@{}} \cite{liu2020general,jiang2019degenerate,yuan2019improving,rosenfeld2017predicting,bonner2018causal,liu2020general} \\ \cite{yu2020influence,chen2021autodebias} \end{tabular}            \\ \cline{2-7}
                                                                                          & \multicolumn{2}{l|}{Reinforcement learning}                               & \begin{tabular}[c]{@{}l@{}}Intervene in the system \\ with a smarter strategy \\ for long-term benefits\end{tabular} & \begin{tabular}[c]{@{}l@{}}Adaptively balancing \\ exploration-exploitation\end{tabular} & \begin{tabular}[c]{@{}l@{}}Hard to train,\\ Off-policy evaluation\\ is chanllenging\end{tabular}                 &  \begin{tabular}[c]{@{}l@{}}\cite{zhao2019deep,li2010contextual,wang2017interactive,wang2017factorization,zhao2013interactive,zhao2020jointly,chen2018stabilizing} \\ \cite{zhao2018recommendations,zheng2018drn,chen2019large,zhao2018deep,zhao2017deep,wang2018reinforcement,jagerman2019people,chen2019top} \\ \cite{jagerman2019people,mcinerney2020counterfactual,swaminathan2017off,jeunen2020joint}\end{tabular}            \\ \cline{2-7}
                                                                                          & \multicolumn{2}{l|}{Others}                                               & -                                                                                                                    & -                                                                                        & -                                                                                                                &  \cite{sun2019debiasing, sinha2016deconvolving}               \\ \hline
\end{tabular}
\label{tb:me}
\end{table*}

\section{Debiasing Methods}
A large number of methods have been proposed  to mitigate the effects of bias or unfairness. Table \ref{tb:me} lists the reviewed methods, as well as their strengths and weaknesses. we classify them according to which biases they addressed and which types of methods they adopted.

\subsection{Methods for Selection Bias}
Training and testing a recommendation model on the observed rating data will suffer from the selection bias, as the observed ratings are not a representative sample of all ratings. Here we fist introduce how to evaluate a recommendation model under biased rating data, and then review four kinds of methods that mitigates selection bias on recommender training.

\subsubsection{Debiasing in evaluation}
 Given a recommendation model, we want to evaluate its performance on rating prediction or recommendation accuracy. Standard evaluation metrics like Mean Absolute Error (MAE), Mean Squared Error (MSE), Discounted Cumulative Gain@k (DCG@k) or Precision (Pre@k) can be written as\cite{schnabel2016recommendations}:
\begin{align}
H(\hat{R})=\frac{1}{|\mathcal U||\mathcal I|} \sum_{u\in \mathcal U} \sum_{i \in \mathcal I} \delta(\hat r_{ui}, {r}_{ui})
\end{align}
for an appropriately chosen $\delta(r_{ui}, \hat{r}_{ui})$:
\begin{align}
\text { MAE: } & \delta(r_{ui}, \hat{r}_{ui})=\left|r_{ui}-\hat{r}_{ui}\right| \\
\text { MSE: } & \delta(r_{ui}, \hat{r}_{ui})=\left(r_{ui}-\hat{r}_{ui}\right)^{2} \\
\text {DCG@k}: \delta(r_{ui}, \hat{r}_{ui}) &=\left(1 / \log \left(\operatorname{rank}\left(\hat{r}_{ui}\right)\right)\right) r_{ui} \\
\text{Pre@k}: \delta(r_{ui}, \hat{r}_{ui}) &=(1 / k) r_{ui} \cdot \mathbf{I}[\operatorname{rank}\left(\hat{r}_{ui}\right) \leq k]
\end{align}
where $\mathbf I[.]$ denotes indicator function ($\mathbf{I}[.]=1$ iff the internal condition holds), $r_{ui}$ denotes the true rating values of the item $i$ given by the user $u$ and $\hat r_{ui}$ denotes the predicted rating values by the recommendation model. As true ratings $r$ are usually partially observed \footnote{Recent work on selection bias usually assumes the conditional distribution $p(R|U,I)$ is stable (\ie $P_E(R|U,I)=P_T(R|U,I)$). Therefore, observed rating values can be considered as true ones $r_{ui}=r^o_{ui}$.}. The conventional evaluation usually use the average over only the observed entries:
\begin{align}
\hat{H}_{\text {naive}}(\hat{r})=\frac{1}{\left|\left\{(u, i): s_{ui}=1\right\}\right|} \sum_{(u, i): s_{ui}=1} \delta(\hat r_{ui}, {r}_{ui})
\end{align}
where $s_{ui}$ denotes the number of observed ratings in the dataset. We can find $\hat{H}_{\text {naive}}(\hat{r})$ is not an unbiased estimate of the true performance \cite{schnabel2016recommendations}:
\begin{align}
E_O\left[ {{{\hat H}_{{\rm{naive}}}}(\hat r)} \right] \ne H(\hat r)
\end{align}
where ${{{\hat H}_{{\rm{naive}}}}(\hat r)}$ is expected over the observation probability. The gap is caused by selection bias, making the observed ratings  not a representative sample of all ratings. Two strategies have been presented in recent work.

\textbf{Propensity Score.} To remedy the selection bias in evaluation, some recent work \cite{schnabel2016recommendations} considers a recommendation as an intervention analogous to treating a patient with a specific drug. In both tasks, we have only partial knowledge of how much certain patients (users) benefit from certain treatments (items), while the outcomes for most patient-treatment (user-item) pairs are unobserved. A promising strategy for both tasks is weighting the observations with inverse propensity scores. The propensity $\rho _{ui}$, which is defined as the marginal probability of observing a rating value ($\rho _{ui}=p(s_{ui}=1)$) for certain user-item pair $(u,i)$, can offset the selection bias. The proposed estimator is defined as:
\begin{align}
{{\hat H}_{IPS}}(\hat r\mid \rho)  =\frac{1}{|\mathcal U||\mathcal I|}\sum\limits_{(u,i):{s_{ui}} = 1} {\frac{\delta(\hat r_{ui}, {r}_{ui})}{{{\rho_{ui}}}}}
\end{align}
which is an unbiased estimator of the ideal metric:
\begin{equation}
\begin{aligned}\small
\mathbb{E}_{S}\left[\hat{H}_{I P S}(\hat{r} \mid \rho)\right] &=\frac{1}{|\mathcal U||\mathcal I|} \sum_{u\in \mathcal U}  \sum_{i\in \mathcal I} \mathbb{E}_{S}\left[\frac{\delta(\hat r_{ui}, {r}_{ui})}{\rho_{ui}} s_{ui}\right] \\
&=\frac{1}{|\mathcal U||\mathcal I|} \sum_{u\in \mathcal U}  \sum_{i\in \mathcal I}  \delta(\hat r_{ui}, {r}_{ui})=H(\hat{r})
\end{aligned}
\end{equation}

\textbf{ATOP.} Steck \etal~ \cite{steck2010training} propose another unbiased metric ATOP to evaluate recommendation performance with two mild assumptions: (1) the relevant (high) rating values are missing at random in the observed data; (2) Concerning other rating values, we allow for an arbitrary missing data mechanism, as long as they are missing with a higher probability than the relevant rating values. They define the ATOP as:
\begin{align}
{\rm{TOPK}}_u^{{\rm{obs}}}(k) &= \frac{{N_u^{ + ,{\rm{obs}},k}}}{{N_u^{ + ,{\rm{obs}}}}}\\
{\rm{TOP}}{{\rm{K}}^{{\rm{obs}}}}(k) & = \mathop \sum \limits_u {w^u}{\rm{TOPK}}_u^{{\rm{obs}}}(k)\\
{\rm{ATO}}{{\rm{P}}^{{\rm{obs}}}} &= \int _0^1{\rm{TOP}}{{\rm{K}}^{{\rm{obs}}}}(k)dk
\end{align}
which computed from biased explicit feedback data and $N_{u}^{+, \mathrm{obs}}$ denotes the number of observed relevant (preferred) items of the user $u$ and $N_{u}^{+, \mathrm{obs},k}$ counts the relevant ones in the top $k$. The authors prove $\mathrm{ATOP}_{u}^{\mathrm{obs}}$ is an unbiased estimate of the average recall and proportional to the precision averaged over users.

\textbf{Discussion.} Propensity scores and ATOP are two subtle strategies to remedy selection bias, but they still have two serve weaknesses. The unbiasedness of the IPS-based estimator is guaranteed only when the true propensities are available\cite{saito2020asymmetric}. The IPS estimator will still be biased if the propensities are specified unproperly. The unbiasedness of the ATOP is guaranteed only when the two assumptions hold. In practice, the missing mechanism is often complex and the assumptions are not always valid. Developing a robust and effective remains a challenge.

\subsubsection{Debiasing in model training} In the following, we will review four kinds of methods on mitigating selection bias on recommender training.

\textbf{Joint Generative Model.} Note that the main reason for the selection bias is that users are free to deliberately choose which items to rate. Thus, a straightforward strategy for mitigating selection bias is to jointly consider both rating prediction task (`which rating value the user gives', \ie $r_{ui}$) and missing data prediction task ('which items the user select to rate', \ie $s_{ui}$). Some recent work \cite{hernandez2014probabilistic,marlin2007collaborative,marlin2009collaborative,jiawei2018social,kim2014bayesian} propose to jointly model the generative process of rating values and the missing mechanism. Their generative process can be depicted with Figure \ref{fg:cabias}(a), with the assumption that the probability of users' selection on items (\ie $s_{ui}$) depends on users' rating values for that item $r_{ui}$. Correspondingly, $s_{ui}$ has been modeled dependent on $r_{ui}$ with a mixture of Multinomials\cite{marlin2007collaborative}, Logit model\cite{marlin2009collaborative,yang2015boosting}, MF model \cite{hernandez2014probabilistic,wang2018deconfounded}, binomial mixture model \cite{kim2014bayesian}, or social-enhanced model \cite{jiawei2018social}. In this way, user's preference can not only learn from rating values but also from the missing mechanism.

Although this kinds of methods is explainable and sometimes effective in some scenarios, jointly modeling the missing mechanism and rating values will lead to a highly complex model, which is hard to be trained. What's worse, the architecture of missing data models are usually heuristically designed. The hypothesis on distribution may not hold in some real cases.


\textbf{Data Imputation.} Note that the inherent nature of selection bias is that the data is missing not random. A straightforward solution for selection bias is to impute the missing entries with pseudo-labels, such that the observed data distribution $p(U,I|S=1)$ is close to the ideal uniform one $P(U,I)$. For example,
Steck \etal~ \cite{steck2013evaluation,steck2010training} propose a light imputation strategy that directly impute the missing data with a specific value $r_0$, with optimizing the following objective function:
\begin{align}
\hat L_{DI}= \sum_{ u\in \mathcal U, i\in \mathcal I} W_{ui} \cdot\left(r_{ui}^{ {o\&i}}-\hat{r}_{ui}\right)^{2}
\end{align}
where $r_{u, i}^{ {o\&i}}$ denotes observed or imputed ratings, while $\hat{r}_{u, i}$ denotes the predicted ratings. $W_{ui}$ is introduced to downweight the contribution of the missing ratings.

However, as imputed rating values are specified in a heuristic manner, this kind of methods will suffer from empirical inaccuracy due to inaccurate imputed rating values. Such inaccuracy will be propagated into recommendation model training, resulting in sub-optimal recommendation performance\cite{wang2019doubly}.

To resolve this issue, Saito \etal~ \cite{saito2020asymmetric} propose to learn imputation values with an asymmetric tri-training framework. They first pre-train two predictors (A1,A2) with two specific recommendation models to generate a reliable dataset with pseudo-ratings and then trained a target recommendation model A0 on the pseudo-ratings. Theoretical analysis presented in \cite{saito2020asymmetric} shows that the proposed method optimizes the upper bound of the ideal loss function. However, the performance of asymmetric tri-training depends on the quality of pre-trained predictor A2, while a satisfied A2 itself is hard obtained from biased data. Nevertheless, model-based imputation strategy is a promising direction for mitigating selection bias, which deserve future exploring.

\textbf{Propensity Score.} Besides on evaluation, propensity score can be utilized to mitigate selection bias on model training \cite{schnabel2016recommendations, wang2021combating}. This kind of methods directly use the IPS-based unbiased estimator as the objective and optimize the following risk function:
\begin{align}
{{\hat L}_{IPS}}= \frac{1}{{|\mathcal U||\mathcal I|}}\sum\limits_{{(u,i):s_{ui}} = 1} {\frac{{{\delta}\left( {\hat r_{ui},r^o_{ui}} \right)}}{{{\rho_{ui}}}}}
\end{align}
Except for the propensities $\rho_{ui}= p(s_{ui}=1)$ that act like weights for each loss term, the training objective is identical to the standard recommendation objective. Also, thanks to the propensities, the selection bias can be mitigated as the IPS-based estimator is an unbiased estimation of the True Risk:
\begin{align}
{\mathbb E}[{{\hat L}_{IPS}}]{\rm{ = }}{\mathbb E_{{S,R}}}[\frac{1}{{|{\mathcal U}||{\mathcal I}|}}\sum\limits_{{u\in \mathcal U, i\in \mathcal I}} {\frac{{s_{ui}\delta \left( {{{\hat r}_{ui}},r_{ui}^o} \right)}}{{{\rho _{ui}}}}} ]{\rm{ = }}\frac{1}{{|{\mathcal U}||{\mathcal I}|}}\sum\limits_{u\in \mathcal U, i\in \mathcal I}{{\mathbb E_{r_{ui}^o \sim {p_E}(R|U,I)}}[\delta \left( {{{\hat r}_{ui}},r_{ui}^o} \right)]} = L
\end{align}

However, as discussed in the previous subsection, specifying appropriate propensity scores is critical. The performance of IPS-based model depends on the accuracy of the propensities. Moreover, propensity-based methods usually suffer from high variance\cite{saito2020asymmetric}, leading to non-optimal results especially when the item popularity or user activeness is highly skewed.


\textbf{Doubly Robust Model.} As data imputation-based models often have a large bias due to mis-specification while IPS-based model usually suffer from high variance, Wang \etal~ \cite{wang2019doubly} propose to combine the two kinds of models and enjoy a desired double robustness property: the capability to remain unbiased if either the imputed errors or propensities are accurate. They define the following objective function:
\begin{align}
{{\mathcal E}_{{\rm{DR}}}}  = \frac{1}{{|{\mathcal U}||{\mathcal I}|}}\sum\limits_{u\in \mathcal U,i \in \mathcal I} {\left( {{{\delta}\left( {\hat r_{ui},r^i_{ui}} \right)} + \frac{{{s_{ui}}({{{\delta}\left( {\hat r_{ui},r^i_{ui}} \right)}-{{\delta}\left( {\hat r_{ui},r^o_{ui}} \right)}})}}{{{{\rho}_{ui}}}}} \right)}
\end{align}
where $r^i_{ui}$ denotes the imputed value for certain user-item pair $(u,i)$. The theoretical and empirical analyses presented in \cite{wang2019doubly} validate superiority over both IPS-based and imputation-based models.

Although the model is more robust than single method, it still requires relatively accurate propensity score or imputation data, which is usually hard to specify. Otherwise, its performance also suffers.

\subsection{Methods for Conformity Bias}
Conformity bias occurs as users are normally influenced by others opinion so that the rating values are deviated from users' true preference. Two types of methods have been proposed to address the conformity bias. The first type of work considers users' behaviors conform to public opinions. For example, Liu \etal~ \cite{liu2016you} directly leverage three important features $c_{ui}, a_{ui}, d_{ui}$ in the base recommendation model, where $c_{ui}$ is the number of ratings for item $i$ before user $u$ rates it, $a_{ui}$ is the average rating and $d_{ui}$ is the rating distribution. The predicted rating is generated from XGBoost \cite{chen2016xgboost}:
\begin{align}
{{\hat r}_{ui}} = xgb\left( {\left\{ {(1 - \omega ) \cdot {t_{ui}} + \omega  \cdot {a_{ui}},{c_{ui}},{a_{ui}},{d_{ui}}} \right\},{\Theta _{xgb}}} \right)
\end{align}
where ${t_{ui}}$ denotes the prediction returned by basic recommendation model and $\omega$ controls the strength of conformity. This way, we can disentangle the effect caused by conformity bias from users' true preference and make a recommendation accordingly. Some recent work further study conformity bias in a more fine-grained manner. Zheng \etal \cite{DBLP:journals/corr/abs-2006-11011} propose to model personalized conformity effect as users have different sensitivities to the public opinions; while Zhao \etal~ \cite{zhao2021popularity} model time-aware conformity effect by considering item dynamic popularity.

The other type of methods treat user's rating values as synthetic results of user preference and social influence \cite{ma2009learning,tang2012mtrust,chaney2015probabilistic,wang2017learning}. Thus, similar to \cite{liu2016you}, they directly leverage social factors in the base recommendation model to generate final prediction and introduce specific parameters to control the effect of social conformity bias.

\subsection{Methods for Exposure Bias}
Exposure bias occurs as users are only exposed to a part of items so that unobserved interactive data does not always mean negative signal. Exposure bias will mislead both the model training and evaluation. Here we review the work on correcting exposure bias.

\subsubsection{Debiasing in evaluation }
A straightforward strategy for debiasing in RS evaluation is using the inverse propersity score, which also has been applied to address the selection bias. Yang \etal~ \cite{yang2018unbiased} first illustrate evaluation bias in terms of conventional metrics such as AUC, DCG@k, Recall@k on the implicit feedback data, and leverage the IPS framework to offset the exposure bias. They abstract the ideal recommendation evaluator as:
\begin{align}
R(\hat Z) = \frac{1}{{|{\mathcal U}|}}\sum\limits_{u \in {\mathcal U}} {\frac{1}{{\left| {{{\mathcal S}_u}} \right|}}} \sum\limits_{i \in {{\mathcal G}_u}} c \left( {{{\hat Z}_{ui}}} \right)
\end{align}
where ${\hat Z}_{ui}$ is the predicted ranking of item $i$ for user $u$ returned by the recommendation model and $\mathcal G_u$ denotes the set of all relevant items for user $u$. Function $c(.)$ needs to be adapted for different metrics, such as:
\begin{align}
\text {AUC}: c\left(\hat{Z}_{u i}\right)&=1-\frac{\hat{Z}_{u, i}}{|I|} \\
\text {DCG}: c\left(\hat{Z}_{ui}\right)&=\frac{1}{\log _{2}\left(\hat{Z}_{ui}+1\right)} \\
\text {DCG@k}: c\left(\hat{Z}_{ui}\right)&=\frac{1\left\{\hat{Z}_{ui} \leq k\right\}}{\log _{2}\left(\hat{Z}_{ui}+1\right)} \\
\text {Recall@k}: c\left(\hat{Z}_{ui}\right)&=1\left\{\hat{Z}_{ui} \leq k\right\}
\end{align}
However, due to the exposure bias, only partial preferred items are observed, making the model often be evaluated on the biased implicit feedback as:
\begin{equation}
\begin{aligned}
\hat{R}_{\mathrm{AOA}}(\hat{Z}) &=\frac{1}{|\mathcal{U}|} \sum_{u \in \mathcal{U}} \frac{1}{\left|\mathcal{G}_{u}^{*}\right|} \sum_{i \in \mathcal G_{u}^{*}} c\left(\hat{Z}_{u i}\right) \\ &=\frac{1}{|\mathcal{U}|} \sum_{u \in \mathcal{U}} \frac{1}{\sum_{i \in \mathcal G_{u}} s_{ui}} \sum_{i \in \mathcal G_{u}} c\left(\hat{Z}_{ui}\right) \cdot s_{ui}
 \end{aligned}
 \end{equation}
where $\mathcal G_u^*$ denotes the preferred items that have been exposed to the user $u$. As users usually have biased exposure, the output of AOA evaluator does not conform the true performance, i.e. $\mathbb{E}_{O}\left[\hat{R}_{\mathrm{AOA}}(\hat{Z})\right] \neq R(\hat{Z})$.

To address this problem, similar to the treatment for selection bias in explicit feedback data, Yang \etal~ \cite{yang2018unbiased} propose to weight the each observation with the inverse of its propensity for implicit feedback data. The intuition is to down-weight the commonly observed interactions, while up-weighting the rare ones. Thus, the IPS-based unbiased evaluator is defined as follow:
\begin{equation}
\begin{aligned}
\hat{R}_{\mathrm{IPS}}(\hat{Z} \mid \rho) &=\frac{1}{|\mathcal{U}|} \sum_{u \in \mathcal{U}} \frac{1}{\left|\mathcal{G}_{u}\right|} \sum_{i \in G_{u}^{*}} \frac{c\left(\hat{Z}_{u i}\right)}{\rho_{ui}} \\
&=\frac{1}{|\mathcal{U}|} \sum_{u \in \mathcal{U}} \frac{1}{\left|\mathcal{G}_{u}\right|} \sum_{i \in \mathcal G_{u}} \frac{c\left(\hat{Z}_{u i}\right)}{\rho_{ui}} \cdot s_{u i}
\end{aligned}
\end{equation}
which is unbiased estimator of the ideal metrics, i.e. ${E_O}\left[ {{{\hat R}_{{\rm{IPS}}}}(\hat Z\mid \rho)} \right]{\rm{ = }}R(\hat Z)$.

\subsubsection{Debiasing in model training}
To deal with the exposure bias and extract negative signal from the implicit feedback, a conventional strategy is treating all the unobserved interactions as negative and specify their confidence. The objective function of most such methods can be summarized as follow:
\begin{align}
\hat L_W=\frac{1}{{|{\mathcal U}||{\mathcal I}|}}\sum\limits_{u\in \mathcal U,i \in \mathcal I} {{W_{ui}}} \delta \left( {{{\hat r}_{ui}},{s_{ui}}} \right) \label{eq:lw}
\end{align}
where $s_{ui}$ is a surrogate label indicating whether the interaction between user $u$  and item $i$ is observed or not; $W_{ui}$ denotes the confidence weight, controlling the confidence that the the feedback of user-item pair $(u, i)$ should be predicted as $s_{ui}$. The specification of the confidence weight is critical to the model performance and can be roughly categorized into three types:

\textbf{Heuristic Weighting.} The first is heuristic-based strategy. For example, the classic weighted factorization matrix (WMF) \cite{hu2008collaborative} and dynamic MF \cite{devooght2015dynamic} used a simple heuristic that the un-observed interactions are assigned with a uniform lower weight, i.e., $W_{ui}=1$ for $s_{ui}=1$ and $W_{ui}=c$ ($0<c<1$) for $s_{ui}=0$. The intuition behind this strategy is that unobserved data is relatively unreliable, which can be attributed to dislike or unknown; Some researchers \cite{pan2009mind,Pan2008} specify the confidence with based on user activity level, i.e., $W_{ui}=c_u, c_u \propto \sum_i{s_{ui}}$, as users associate with more items provide more reliable information; Analogously, item popularity has been considered to specify confidence weights \cite{he2016fast,yu2017selection}, as popular items are more probable to be exposed; Also, user-item feature similarity  \cite{li2010improving} has been considered to define the confidence.

However, assigning appropriate confidence weights heuristically is challenging, as the optimal data confidence may change for different user-item combinations. Choosing confidence weights usually require rich human expertise or large computational resource for grid search. Furthermore, it is unrealistic to manually set flexible and diverse weights for millions of data. Coarse-grained confidence weights will create empirical bias on estimating user's preference.

\textbf{Sampling.} Another solution to address exposure bias is performing sampling. The sampling strategy determines which data are used to update parameters and how often, and thus
 scale the data contribution. Provided the sampled probability of an instance is $p_{ui}$, learning a recommendation model with sampling is equivalent to learning the model with the following weighted objective function:
\begin{align}
{\mathbb E_{(u,i)\sim p}}[\delta \left( {{{\hat r}_{ui},{s_{ui}}}} \right)] = \sum\limits_{u\in \mathcal U,i \in \mathcal I} {{p_{ui}}} \delta \left({{\hat r}_{ui}}, {{s_{ui}}} \right)
\end{align}
where the sampling distribution acts as data confidence weights. Sampling strategy has been widely applied as its efficiency. For example, Logistical matrix factorization \cite{johnson2014logistic}, BPR \cite{rendle2009bpr}, or most of neural-based recommendation models (e.g. CDAE \cite{wu2016collaborative}, NCF \cite{he2017neural}, LightGCN \cite{he2020lightgcn}) apply the uniform negative sampler; Yu \etal~ \cite{yu2017selection} considers to over-sample the popular negative items, as they are more likely to be exposed. However, these heuristic samplers are insufficient to capture real negative instances. Thus, some researchers explore to leverage side information to enhance the sampler. Ding \etal~ \cite{ding2018improved,ding2019reinforced}  leverage viewed but non-clicked data to evaluate user's exposure; Chen \etal~ \cite{chen2019samwalker} leverage social network information in their sampling distribution; Wang \etal~ \cite{wang2020reinforced} construct an item-based knowledge graph and perform sampling on the graph.

\textbf{Exposure-based model.} Another strategy is to develop an exposure-based model, which is capable of capturing how likely a user is exposed to an item\cite{DBLP:conf/aaai/WangZYZ18, chen2018modeling}.
EXMF \cite{liang2016modeling} introduces an exposure variable and assumes the following generative process of implicit feedback:
\begin{align}
   {e_{ui}}&\sim Bernoulli({\eta _{ui}}) \hfill \\
  ({s_{ui}}|{e_{ui}} = 1)&\sim Bernoulli(\hat r_{ui}) \hfill \\
  ({s_{ui}}|{e_{ui}} = 0)&\sim {\delta _0}
\end{align}
where $e_{ui}$ denotes whether a user $u$ has been exposed to the item $i$; $\delta _0$ denotes delta function $p(s_{ui}=0|e_{ui}=0)=1$ and can be relaxed as Bernoulli distribution parameterized with a small value; ${\eta _{ui}}$ is the prior probability of exposure.  When $e_{ui}=0$, we have $s_{ui}\approx 0$, since when the user does not know the item he can not interact with it. When $e_{ui}=1$, \ie the user has known the item, he will decide whether or not to choose the item based on his preference. $s_{ui}$ can be generated with normal recommendation model. In this way, by optimizing the marginal probability, the model can adaptively learn the exposure probability, which will be transformed as confidence weights to remedy exposure bias. Chen \etal \cite{chen2019samwalker,chen2020fast} give detailed analyses of EXMF and rewrite the objective function of EXMF as follows:
\begin{align}
\hat L_{EXMF}=\sum\limits_{u\in \mathcal U,i\in \mathcal I} {{\gamma _{ui}}\delta (\hat r_{ui},{x_{ui}})}  + \sum\limits_{u\in \mathcal U,i\in \mathcal I} {g({\gamma _{ui}})} \label{eq:exmf}
\end{align}
where $\gamma _{ui}$ is defined as variational parameters of the user's exposure. $g({\gamma _{ui}})$ is a $\gamma _{ui}$-dependent function:
\begin{align}
g({\gamma _{ui}}) =(1 - {\gamma _{ui}})\ell ({x_{ui}},\varepsilon ) + \ell ({\gamma _{ui}},{\eta_{ui}}) - \ell ({\gamma _{ui}},{\gamma _{ui}})
\end{align}
where $\ell(a,b)=alog(b)+(1-a)log(1-b)$. We can find $\gamma _{ui}$, which indicates how likely a user is exposed to an item, acts as confidence weights to control the contribution of the data on learning a recommendation model. This finding is consistent with our intuition. Only if the user has been exposed to the item, can he decide whether or not to consume the items based on his preference. Thus, the data with larger exposure are more reliable in deriving user preference.


However, directly estimating data confidence from Equation (\ref{eq:exmf}) is insufficient as the model will easily suffer from over-fitting and inefficiency problems due to the large scale of the inferred parameters $\gamma$. A promising solution is to re-parameterize the confidence weights with a simpler function. For example, some researchers propose to infer confidence wights with a social-based \cite{chen2019samwalker} or community-based model \cite{chen2020fast,wang2021samwalker++}.

\textbf{Propensity Score.} Although the aforementioned weighting strategies are popular and have been studied for a long time, Saito \etal~ \cite{saito2020unbiased} argue that these methods can not address exposure bias entirely --- For any choice of the weights ($\forall W_{ui}\in \mathbb R$), The weighted empirical risk $L_W$ can not be an unbiased estimator of the ideal \textit{True Risk}, \ie ${\mathbb E}[{{\hat L}_W}] \ne L$. To tackle this problem, Saito \etal~ \cite{saito2020unbiased} proposes a new estimator with propensity score as follow:
\begin{align}
{{\mathcal {\hat L}}_{{\rm{sur}}}} &= \frac{1}{{|{\mathcal U}||{\mathcal I}|}}\sum\limits_{u\in \mathcal U,i \in \mathcal I} {\left[ {{s_{ui}}\left(\frac{1}{{{\rho_{ui}}}}\delta \left( {{{\hat r}_{ui}},1} \right) + (1 - \frac{1}{{{\rho_{ui}}}})\delta \left( {{{\hat r}_{ui}},0} \right)\right )}
 { + \left( {1 - {s_{ui}}} \right)\delta \left( {{{\hat r}_{ui}},0} \right)} \right]}
\end{align}
where the propensity score $\rho_{ui}=p(e_{ui}=1)$ is defined as the marginal probability of a user exposed to the item. ${{{\hat L}}_{{\rm{sur}}}}$ is an unbiased estimator of the \textit{True Risk}:
\begin{equation}
\begin{aligned}
\mathbb E[{{{\hat L}}_{{\rm{sur}}}}]&=\mathbb E_{r_{ui},e_{ui}}\left[\frac{1}{{|{\mathcal U}||{\mathcal I}|}}\sum\limits_{u\in \mathcal U,i \in \mathcal I} {\left[ {e_{ui}r_{ui} \left(\frac{1}{{{\rho_{ui}}}}\delta \left( {{{\hat r}_{ui}},1} \right) + (1 - \frac{1}{{{\rho_{ui}}}})\delta \left( {{{\hat r}_{ui}},0} \right)\right )}
 { + \left( {1 - e_{ui}r_{ui}} \right)\delta \left( {{{\hat r}_{ui}},0} \right)} \right]}\right] \\
 &=\mathbb E_{r_{ui}}\left[\frac{1}{{|{\mathcal U}||{\mathcal I}|}}\sum\limits_{u\in \mathcal U,i \in \mathcal I} {\left[ r_{ui}\delta \left( {{{\hat r}_{ui}},1} \right)+(1-r_{ui})\delta \left( {{{\hat r}_{ui}},0} \right) \right]}\right]=\mathbb E_{r_{ui}}\left[\frac{1}{{|{\mathcal U}||{\mathcal I}|}}\sum\limits_{u\in \mathcal U,i \in \mathcal I}\delta \left( {{{\hat r}_{ui}},r_{ui}}\right)\right]\\
 &=L
\end{aligned}
\end{equation}

This propensity-based strategy is theoretical soundness and usually achieves better performance than weighting strategies. It is flexible and also has been extended to pair-wise objective function \cite{rendle2009bpr,wang2021non,saito2020unbiased}. However, this kind of methods has some limitations: (1) its performance depends on the accuracy of the propensity score, which is quite challenging to obtain; (2) the inverse of propensity incurs high variance. Although these problems can be mitigated
to a certain extent by some strategies (\eg joint learning \cite{zhu2020unbiased}, clapping \cite{saito2020unbiased}), they still deserve further exploration.

\textbf{Causality-based Methods.} Causal inference is another promising direction for addressing exposure bias. In fact, the spirit of a recommendation can be understood as to answer a counterfactual question: would the user interacts with the item if he had know the item? That is, we need to evaluate the causal estimand $p(S|do(E=1),U,I)$ with intervening item exposure\footnote{Here the do-calculus $do(.)$ indicates the variable $S$ is coercively intervened with a certain value. For more details on causal inference, we refer the readers to the excellent causal textbook \cite{pearl2018book}. } rather than the statistical associations $p(S|U,V)$ estimated by vanilla recommender models. The intervention could remove the spurious association caused by exposure bias and recover users' true preference on the items.

 Towards this target, various causality-based methods have been proposed. For example, Zhang \etal \cite{zhang2021causal} resorted to the back-door criterion \cite{pearl2018book} to remove the exposure bias caused by the item popularity; Xu \etal \cite{xu2021deconfounded} leveraged forward door criterion \cite{pearl2018book} to remove the effect from unobserved confounders; Wang \etal \cite{wang2020click} leveraged counterfactual reasoning to eliminate the direct causal effect from exposure features on the prediction; Yang \etal \cite{yang2021top} mitigates exposure bias through counterfactual samples; Liu \etal \cite{liu2021mitigating} disentangle the effect from exposure and preference with introducing information bottleneck.


\textbf{Others.} There are also some other strategies to address exposure bias in specific scenarios. Wang \etal \cite{wang2016learning} consider the queries of a search system are usually under-sampled to different extents, and thus are biased when click data is collected to learn the ranking function. They further propose a specific model for this situation, where queries are classified into different classes, and the bias in each class is estimated with randomized data. Ovaisi \etal \cite{ovaisi2020correcting,DBLP:conf/sigir/OvaisiVZ21} attribute exposure bias to the fact that a user can examine only a truncated list of top-K recommended items. To address this kind of exposure bias, two-step Hechman method has been adopted. They first use a Probit model to estimate the probability of a document being observed and then leverage the exposure probability to correct the click model. Some recent work also consider users' sequential behaviors ``exposure-click-conversion'' and correspondingly devise an inverse propensity model \cite{zhang2020large}, decomposition model \cite{ma2018entire,wen2020entire} or graph neural network \cite{bao2020gmcm} on the sequential behaviors to address exposure bias with multi-task learning. Besides, propensity scoring model has been utilized in debiasing explainable recommendation \cite{damak2021debiased} and item-to-item recommendation \cite{schnabel2020debiasing}.

\subsection{Methods for Position Bias}
Position bias is another type of bias that is widely studied in learning-to-rank systems, such as ad system and search engine. Position bias denotes that the higher ranked items will be more likely to be selected regardless of the relevance. Recent years have seen a number of work on position bias, and we categorize them into three lines.

\textbf{Click Models.}
The first line is based on click models. The methods make hypotheses about user browsing behaviors and estimate true relevance feedback by optimizing the likelihood of the observed  clicks. some work \cite{craswell2008experimental,dupret2008user,chapelle2009dynamic,zhang2007comparing,lin2021graph} on click models assume the examination hypothesis that if a displayed item is clicked, it must be both examined and relevant. This is based on the eye-tracking studies which testify that users are less likely to click items in lower ranks. To remedy position bias and to recover user true preference, they explicitly model the probability of an user clicks an item $i$ at position $q$ as follows:
\begin{equation}
\begin{aligned}
&{P(C = 1\mid u,i,p)}\\
&{ = \underbrace {P(C = 1\mid u,i,E = 1)}_{{r_{ui}}} \cdot \underbrace {P(E = 1\mid q)}_{{h_q}}}
\end{aligned}
\end{equation}
Notice that a hidden random variable $E$ has been applied, which denotes whether the user has examined the item. In general, these methods make the following assumptions: if the user clicks it, the item must have been examined; if the user has examined the item, the click probability only depends on the relevance; and the examination depends solely on the position $p$. The model is highly similar to the exposure-based model for exposure bias except that the exposure probability is modeled with position.

Another choice of click model is the cascade model \cite{craswell2008experimental}. It differs from the above model in that it aggregates the clicks and skips in a single query session into a single model. It assumes a user examines an item from the first one to the last one, and the click depends on the relevance of all the items shown above. Let $E_q, C_q$ be the probabilistic events indicating whether the $q$-th item is examined and clicked respectively. The cascade model generates users click data as follows:
\begin{align}
&P\left( {{E_1}} \right) = 1\\
&P\left( {{E_{q + 1}} = 1\mid {E_q} = 0} \right) = 0 \\
&P\left( {{E_{q + 1}} = 1\mid {E_q} = 1,{C_q}} \right) = 1 - {C_q} \label{eq:33} \\
&P\left( {{C_q} = 1\mid {E_q} = 1} \right) = {r_{{u_q},i}}
\end{align}
in which the Equation (\ref{eq:33}) implies that if a user finds her desired item, she immediately closes the session; otherwise she always continues the examination. The cascade model assumes that there is no more than one click in each query session, and if examined, an item is clicked with probability ${r_{{u_q},i}}$ and skipped with $1-{r_{{u_q},i}}$. This basic cascade model has been further improved by considering the personalized transition probability  \cite{chapelle2009dynamic,guo2009click,zhu2010novel}. Jin \etal~ \cite{jin2020deep} improve these models and consider users browsing behaviors in a more thorough manner with deep recurrent survival model.

However, these click models usually require a large quantity of clicks for each query-item or user-item pair, making them difficult to be applied in systems where click data is highly sparse, e.g., personal search \cite{wang2016learning}. Further, mis-specifying the generative process of users clicks will cause empirical bias and hurt recommendation performance.

\textbf{Propensity Score.}
Another common solution to correct position bias is employing inverse propensity score, where each instance is weighted with a position-aware values \cite{agarwal2019general}. The loss function is defined as follow:
\begin{equation}
\begin{aligned}
{L_{{\rm{IPS}}}}(f){\rm{ = }}\sum\limits_{u\in \Set U,i \in \Set I} {\frac{1}{{\rho (q)}}{s_{ui}}\lambda (u,i|f)}
\end{aligned}
\end{equation}
Here we refer to \cite{vardasbi2020inverse} and use the ranking metrics. $\lambda (u,i|f)$ denotes the metric function that is based on the the rank of the item $i$ for the user (or query) $u$ according to the ranking system $f$. For instance, it can be chosen to match the well-known NDCG metric:
\begin{equation}
\begin{aligned}
{\lambda _{NDCG}}(u,i|f) = {\left( {{{\log }_2}\left( {{\mathop{\rm rank}\nolimits} \left( {i|u,f} \right) + 1} \right)} \right)^{ - 1}}
\end{aligned}
\end{equation}
A position-dependent propensity $\rho(q)$ is introduced to weight the $\lambda$ function. The intuition behind the model is that clicks on items that are less likely to have been examined by users are weighted more heavily. This weighting strategy compensates for the effect of position bias on user exposure, allowing the method to estimate and learn without being affected by position bias in expectation Joachims \etal~ \cite{joachims2017unbiased}.

Estimating the propensity score for position bias have been well explored as its simplicity --- just dependent on the item position. A simple yet effective solution to estimate a position-based propensity model is result randomization, where the ranking results are shuffled randomly and collect user clicks on different positions to compute propensities scores \cite{schuth2016multileave,wang2016learning,raman2013learning,swaminathan2015batch,hofmann2013reusing}. Because the expected item relevance is the same on all positions, it is provable that the difference of click rate on different positions produces an unbiased estimation of the truth propensities. Despite its simplicity and effectiveness, result randomization has a risk of significantly hurting the user experience as the highly ranking items may not be favored by the user. Pair-wise swapping \cite{joachims2017unbiased} has been proposed to mitigate the problem, but can not eliminate negative effect completely. Therefore, the strategies that learn the propensity scores from the data without any intervention on the recommendation results have been explored. Fang \etal~ \cite{fang2019intervention} and Agarwal \etal~ \cite{agarwal2019estimating} adopt intervention harvesting, to learn the propensity. However, such methods require the feedback data from multiple ranking models. Further, some recent work \cite{ai2018unbiased,wang2018position,joachims2017unbiased,qin2020attribute} consider learning a propensity model and a recommendation model as dual problem and develop specific EM algorithms to learn both models. More recently, the click model that captures the row skipping and slower decay phenomenon has been adopted to specify the propensity scores in \cite{guo2020debiasing}, while cascade model has been adopted by \cite{vardasbi2020cascade}. Chen also \etal \cite{chen2021adapting} propose to learn the propensity from the data observation.

\textbf{Trust-aware Models.} Item position not only influences users' exposure but also their decisions (\ie Both $R$ and $E$ are dependent on $Q$). Aforementioned propensity score is insufficient to address this problem. Hence, Agarwal \etal \cite{agarwal2019addressing} propose an expansion to IPS to correct for both effects. The model hypothesizes that a real relevant item at position $q$ can be misjudged with probability $1-\epsilon_{q}^{+}$, while a non-relevant item can be clicked mistakenly with probability $\epsilon_{q}^{-}$, \ie we have:
\begin{equation}
\begin{aligned}
p(R = 1|U,I,Q) &= p({r_{ui}}=1)\epsilon_{q}^{+}+p({r_{ui}}=0)\epsilon_{q}^{-} \\
p(R = 0|U,I,Q) &= p({r_{ui}}=1)(1-\epsilon_{q}^{+})+p({r_{ui}}=0)(1-\epsilon_{q}^{-})
\end{aligned}
\end{equation}
To tackle the label inversion caused by position bias, Agarwal \etal \cite{agarwal2019addressing} extend IPS to the following objective:
\begin{equation}
\begin{aligned}
{L_{{\rm{Bayes-IPS}}}}{\rm{(}}f){\rm{ = }}\sum\limits_{u,i} {\frac{\epsilon_{k}^{+}}{\epsilon_{k}^{+}+\epsilon_{k}^{-}}\frac{1}{{\rho (q)}}{s_{ui}}\lambda (u,i|f)}
\end{aligned}
\end{equation}
where an offset term ${\epsilon_{k}^{+}}/({\epsilon_{k}^{+}+\epsilon_{k}^{-}})$ is introduced.  Vardasbi \etal \cite{vardasbi2020inverse} further proof that ${L_{{\rm{Bayes-IPS}}}}$ is insufficient and propose a more theoretical-soundness method with affinity corrections:
\begin{equation}
\begin{aligned}
{L_{{\rm{Affinity}}}}{\rm{(}}f){\rm{ = }}\sum\limits_{u,i} {\frac{{{s_{ui}}}-\rho (q) \epsilon_{k}^{-}}{\rho (q)\left(\epsilon_{k}^{+}-\epsilon_{k}^{-}\right)}\lambda (u,i|f)}
\end{aligned}
\end{equation}
which is an unbiased estimation of the ideal estimator \wrt position bias.

\subsection{Universal Solutions for Various Data Biases}
Most existing methods are designed for addressing one or two biases of a specific scenario. Hence, when facing the real data that commonly contain multiple types of biases, these methods will fall short. Recently saw a few studies on universal solutions for multiple data biases and their combinations. These methods resorted to a small unbiased dataset for recommendation debiasing. For example, some work transfered the knowledge from the unbiased data to the target model with domain adaption \cite{bonner2018causal,lin2021transfer} or knowledge distillation \cite{liu2020general}; More recently, Chen \etal \cite{chen2021autodebias} proposed to learn the optimal debiasing configures from the uniform data with meta learning.

Despite their effectiveness on handle various data biases, these methods require unbiased data, which is difficult and expensive to obtain. To collect uniform data, we must intervene in the system by using a random logging policy instead of the normal recommendation policy, which would hurt users' experience and revenues of the platform. Therefore, how to develop a universal solution without using unbiased data is still an open problem and deserves further exploration.

\subsection{Methods for Popularity Bias}

Popularity bias is a common problem in recommendation systems. We categorize the methods into four types.

\textbf{Regularization.}
Suitable regularization can push the model towards balanced recommendation lists. Abdollahpouri \etal~ \cite{DBLP:conf/recsys/AbdollahpouriBM17} introduced \textit{LapDQ} regularizer~\cite{DBLP:conf/flairs/WasilewskiH16} $tr(Q^T L_DQ)$, where $Q$ denotes the item embedding matrix, $tr(\cdot)$ denotes the trace of a matrix, and $L_D$ denotes the Laplacian matrix of $D$, where $D_{i,j}=1$ if item $i$ and $j$ belong to the same set (popular items or long-tail items) and $0$ otherwise.
Kamishima \etal~ \cite{DBLP:conf/recsys/KamishimaAAS14} applied the \textit{mean-match} regularizer ~\cite{kamishima2013efficiency} in their \textit{information-neutral recommender systems (INRS)} to correct popularity bias.
They first introduced mutual information to measure the influence of features on the recommendation results,
and through a series of mathematical approximations and derivations, they obtain a specific regularization term: $-(\bm M_{D^{(0)}}(\{\hat r\})-\bm M_{D^{(1)}}(\{\hat r\}))^2$,
where $\bm M_D (\{\hat r\}) = \frac{1}{|D|} \sum_{(x_i,y_i,v_i)\in D} \hat r(x_i, y_i, v_i)$. More recently, Zhu \etal~ \cite{zhu2021popularity} utilized a \textit{Pearson Coefficient} regularizer to decrease the correlation between item popularity and model output scores. Note that the above regularizers are result-oriented, guiding the model to give more balanced results.

Different from result-oriented regularizers, Chen \etal~ \cite{DBLP:conf/sigir/ChenXLYSD20} devise a process-oriented regularization term.
It attributes the inability of effectively recommending long-tail items as the insufficient training of them. These items usually have few interaction records and thus their embedding vectors can not be well trained, making their prediction scores close to the initial values and remain neutral.
Motivated by this point, Chen \etal~ proposed \textit{Entire Space Adaptation Model (ESAM)} from the perspective of domain adaptation (DA). ESAM aims to transfer the knowledge from these well-trained popular items to the long-tail items.
ESAM introduced three regularization terms for transferring as:
(1) Domain adaptation with item embedding (or attributes) correlation alignment: $L_{DA} = \frac{1}{L^2}
\sum_{(j,k)}({\bm h_s^j}^T \bm h_s^k - {\bm h_t^j}^T \bm h_k^t )^2 = \frac{1}{L^2} || Cov(\bm D^s) - Cov(\bm D^t)  ||_F^2$,
where $|| \cdot||^2_F$ denotes squared matrix Frobenius norm.
$Cov(\bm D^s) \in R^{L*L}$ and $Cov(\bm D^s) \in R^{L*L}$ represent the covariance matrices of
high-level item attributes, which can be computed as $Cov(\bm D^s) ={\bm D^s}^T \bm D^s $,
and $Cov(\bm D^t) ={\bm D^t}^T \bm D^t $. $s$ means source domain (popular items), and $t$ means target domain (unpopular items).
(2) Center-wise clustering for source clustering $L_{DC}^c$: encouraging the features of the items with the same feedback (such as buy, view, and click) to be close together, and the features of the items with different feedbacks to move away from each other.
(3) Self-training for target clustering $L_{DC}^p$: minimizing the entropy regularization $-p logp$ favors a low-density separation between classes. This term is a way of self-training which increases the discriminative power between non-displayed items.

\textbf{Adversarial Learning.}
Adversarial learning is another line to address popularity bias\etal \cite{DBLP:conf/cikm/KrishnanSSS18}. The basic idea is to play a min-max game between the recommender G and the introduced adversary D, so that D gives a signal to improve the recommendation opportunity of the niche items. In \cite{DBLP:conf/cikm/KrishnanSSS18},
The adversary D takes the synthetically generated popular-niche item pairs $(\tilde i^p,\tilde i^n | u)$, and an equal number of true popular-niche pairs $(i^p,i^n)$ as input.
True pairs $(i^p,i^n)$ are sampled from their global co-occurrence and synthetic pairs $(\tilde i^p,\tilde i^n)$ are drawn by the recommender. The recommender G can be instantiated with recent recommendation model such as NCF.
Through adversarial learning between G and D, D learns the implicit association between popular and niche items, while G learns to capture more niche items that correlate with the user's history, resulting in recommending more long-tail items for users.

\textbf{Causal Graphs.}
Causal graph is a powerful tool for counterfactual reasoning. Some recent work proposed to leverage causal graph to tackle popularity bias. They first built a causal graph to elucidate popularity bias, and then applied counterfactual intervention over the graph to mitigate the bias. For example, Zhang \cite{zhang2021causal} \etal~ attributed the popularity bias to the undesirable causal effect from item popularity to the item exposure. To eliminate this effect, they further proposed to intervene the distribution of the exposed items with back-door criterion or propensity score; Zhao \etal~ \cite{zhao2021popularity} and Wang \etal~ \cite{DBLP:journals/corr/abs-2006-11011} traced popularity bias from conformity effect (\ie the effect of item popularity on user behavior), and causally intervened the item popularity to make fair recommendation; Analogically, Wei \etal~ \cite{wei2021model} performed counterfactual reasoning to eliminate the direct effect of item (popularity) to the prediction; Wang \etal~ \cite{DBLP:conf/kdd/WangF0WC21} studied how popularity bias occurs in model training. They attributed the popularity bias to a confounding causal structure and applied backdoor adjustment to mitigate this effect.

%
%
%
%


\textbf{Others.}
There are some other methods on popularity bias. one solution to reduce popularity bias is through introducing other side information. For example, Bressan \etal~ leverage social information to reduce popularity bias~\cite{DBLP:conf/kdd/Bressan0PRT16}. Abdollahpouri gives a different strategy ~\cite{DBLP:conf/aies/Abdollahpouri19}, which relies on re-ranking. To perform top-k recommendation, it first generates a relatively large recommendation list with a classical model, and then re-ranks the list by considering the item popularity. Similar to exposure bias, propensity score can also be applied to reduce popularity bias: by decreasing the influence of popularity items to model training, the popularity bias can be mitigated~\cite{yang2018unbiased}.

\subsection{Methods for Unfairness}
Before introducing existing fairness-aware methods, we first give some formulations of fairness.

\subsubsection{Fairness Formulations}
There are extensive studies on fairness in machine learning. Without loss of generality, we use the notation of prediction model throughout this section to discuss fairness. Let $A$, $X$, $U$ be the set of sensitive attributes (\aka protected attributes), other observed attributes, and unobserved attributes of an individual, respectively. $Y$ denotes the ground-truth outcome to be predicted, while $\hat{Y}$ is the prediction produced by a prediction model that depends on $A$, $X$, $U$. For simplicity we often assume A is encoded as a binary attribute, but this can be generalized to other cases.

There exist many different variations of fairness definition, which can be roughly categorized into four types: 1) \textbf{fairness through unawareness}~\cite{grgic2016case}; 2) \textbf{individual fairness}~\cite{DBLP:conf/innovations/DworkHPRZ12,DBLP:journals/corr/JosephKMNR16,DBLP:journals/corr/LouizosSLWZ15,LFR}; 3) \textbf{group fairness} (\eg demographic parity~\cite{DBLP:conf/aistats/ZafarVGG17,kimfact}, equality of opportunity~\cite{DBLP:conf/nips/HardtPNS16,DBLP:conf/www/ZafarVGG17}, predictive equality~\cite{DBLP:journals/bigdata/Chouldechova17}, equalized odds~\cite{DBLP:conf/nips/HardtPNS16}, calibration within groups~\cite{DBLP:conf/innovations/KleinbergMR17}); and 4) \textbf{counterfactual fairness}~\cite{Counterfactual-Fairness}.
Here we present some widely-used formulations:
\begin{itemize}
    \item \textbf{Fairness Through Unawareness}: \emph{A model is fair if any sensitive attributes $A$ are not explicitly used in the modeling process.}
    \item \textbf{Individual Fairness}: \emph{A model is fair if it gives similar predictions to similar individuals. Formally, if individuals $i$ and $j$ are similar under a certain metric, their predictions should be similar: $\hat{Y}(X(i), A(i))\approx \hat{Y}(X(j), A(j))$.}
    \item \textbf{Demographic Parity}: \emph{Each protected group (\ie with the same sensitive attributes) should receive positive prediction at an equal rate. Formally, the prediction $\hat{Y}$ satisfies demographic parity if $P(\hat{Y}|A=0)=P(\hat{Y}|A=1)$.}
    \item \textbf{Equality of Opportunity}: \emph{Given the prediction model, the likelihood of being in the positive class is the same for each protected group. Formally, the prediction $\hat{Y}$ satisfies the equality of opportunity if $P(\hat{Y}=1|A=0, Y=1) = P(\hat{Y}=1|A=1, Y=1)$.}
    \item \textbf{Counterfactual fairness}: \emph{Given a causal model $(U, A\cup X, F)$, the prediction $\hat{Y}$ is counterfactually fair if under any context $X = x$ and $A = a$, $P(\hat{Y}_{A\leftarrow a}(U) = y | X = x, A = a) = P(\hat{Y}_{A\leftarrow a'}(U) = y | X = x, A = a)$, for all $y$ and for any value $a'$ attainable by $A$.}
\end{itemize}
Besides these general definitions \wrt user attributes, the concept of fairness has been generalized to multiple dimensions in recommender systems~\cite{DBLP:journals/corr/Burke17aa}, spanning from fairness-aware ranking~\cite{DBLP:conf/kdd/GeyikAK19,DBLP:conf/sigmod/AsudehJS019,DBLP:conf/sigir/BiegaGW18}, fairness in terms of user psychological characteristics \cite{wang2021user}, supplier fairness in two-sided marketplace platforms~\cite{DBLP:conf/cikm/MehrotraMBL018}, provider-side fairness to make items from different providers have a fair chance of being recommended~\cite{DBLP:conf/recsys/KamishimaAAS14,DBLP:conf/recsys/LiuGSBZ19}, fairness in group recommendation to minimize the unfairness between group members~\cite{DBLP:conf/recsys/LinZZGLM17}.

\subsubsection{Fairness-aware Methods}
In the following, we review four different ways to mitigate the unfairness issue on recommendation.

\textbf{Rebalancing.}
Inspired by the strategy used to tackle the class-imbalance problem, one common paradigm is to balance the data or recommendation results \wrt certain fairness target like demographic parity. Some intensively-adopted strategies in machine learning research are re-labeling the training data to make the proportion of positive labels equal in the protected and unprotected groups~\cite{DBLP:conf/kdd/PedreschiRT08}, or re-sampling the training data to achieve statistical parity~\cite{DBLP:conf/kdd/GeyikAK19}.

This idea of rebalancing data is prevalent in fairness-aware ranking, where the fairness constraint can be represented in various forms. Towards individual equity-to-attention fairness in rankings, previous work~\cite{DBLP:conf/sigmod/AsudehJS019,DBLP:conf/sigir/BiegaGW18} propose multiple ranking functions to sort items and then achieve fairness amortized across these rankings.
Towards group fairness, FA$*$IR~\cite{DBLP:conf/cikm/ZehlikeB0HMB17} is a post-processing method to achieve fair top-$K$ ranking \wrt group fairness criteria, in which a subset of $K$ candidates are re-selected from a large item collection to achieve a required proportion for a single under-represented group.
Analogously, DetCons and DetConstSort~\cite{DBLP:conf/kdd/GeyikAK19} formalize the fairness as a desired distribution over sensitive attributes, and re-rank candidates (\ie LinkedIn users) to satisfy the constraints; Li \etal~ \cite{li2021user} constrain the difference of the average recommendation performance between two groups, and formulate the fairness-aware ranking problem as 0-1 integer programming.
To formulate group fairness in terms of exposure allocation, Singh~\etal~\cite{DBLP:conf/kdd/SinghJ18} propose a framework for formulating fairness constraints on rankings, and sample rankings from an associated probabilistic algorithm to fulfill the constraints. HyPER~\cite{DBLP:journals/corr/abs-1809-09030} uses probabilistic soft logic (PSL) rules to balance the ratings for both users in protected and unprotected groups, where fairness constraints are encoded as a set of rules.
More recently, when organizing user-item interactions in the form of graph, some work~\cite{Fairwalk,DBLP:journals/corr/abs-2002-11442} study potential unfairness issue inherent within graph embedding. Among them, Fairwalk~\cite{Fairwalk} treats the group information \wrt sensitive attributes as a prior distribution, and then performs node2vec based on the prior to sample random walks and generate debiased embeddings, which are evaluated in friendship recommendation.

\textbf{Regularization.}
The basic idea of the regularization line is to formulate the fairness criteria as a regularizer to guide the optimization of model.
A general framework, \emph{Learned Fair Representation (LFR)}, is proposed in~\cite{LFR}, which generates the data representations to encode insensitive attributes of data, while simultaneously removing  any information about sensitive attributes \wrt the protected subgroup.
Formally, it is composed of three loss components:
\begin{gather}\label{equ:fairness-lfr}
    \min\Lapl = \alpha C(X,R) + \beta D(R,A) + \gamma E(Y,R)
\end{gather}
where $C(\cdot)$ is the reconstruction loss between input data $X$ and representations $R=Enc(X)$ with an encoder function $Enc(\cdot)$; $E(\cdot)$ is the prediction error in generating prediction $Y$ from $R$, such as cross entropy; $D(\cdot)$ is a regularization term that measures the dependence between $R$ and sensitive attribute $A$, which is defined as fairness constraints such as demographic parity:
\begin{gather}
    D(R,A) = |\Space{E}_{R}P(R|A=1)-\Space{E}_{R}P(R|A=0)|
\end{gather}
where $P(R|A=1)$ relies on the distance of representation $R$ and the centroid representation $\tilde{R}_{1}$ of the group where $A=1$:
\begin{gather}
    P(R|A=1) = \frac{\exp{-||R-\tilde{R}_{1}||_{2}}}{\sum_{a\in\{0,1\}}\exp{-||R-\tilde{R}_{a}||_{2}}}
\end{gather}
Using such a regularization makes the encoded representation sanitized and blind to whether or not the individual $X$ is from the protected group.

Studies on this research line have been extensively conducted by subsuming different fairness formulations under the foregoing framework.
Earlier, Kamishima~\etal~first claimed the importance of neutrality (\aka viewpoint of sensitive attribute) in recommendation~\cite{DBLP:conf/recsys/KamishimaAAS12}, and then proposed two methods --- (1) one regularization-based matrix completion method~\cite{kamishima2013efficiency}, where the fairness regularizer is formulated as the negative mutual information $-I(A;Y)$ between sensitive attribute $A$ and prediction $Y$, and (2) one graphical model-based method~\cite{DBLP:conf/icdm/KamishimaAAS16}, where the fairness regularizer accounts for the expected degree of independence between $A$ and $Y$ in the graphical model.
Later, Kamishima~\etal~generalized these work to implicit feedback-based recommender systems~\cite{kamishima2017considerations}.
Analogously, Yao~\etal~\cite{DBLP:conf/nips/YaoH17,DBLP:journals/corr/YaoH17a} proposed four fairness metrics in collaborative filtering, and used similar regularization-based optimization method to mitigate different forms of bias.

Moreover, there are some regularization-based studies working on more specific scenarios.
For example, Abdollahpouri \etal~\cite{DBLP:conf/recsys/AbdollahpouriBM17} focused on controlling popularity bias in learning-to-rank recommendation, and proposed a regularizer that measures the lack of fairness for the short-head and medium-tail item sets in a given recommendation list to improve fairness during model training.
Xiao \etal~\cite{DBLP:conf/recsys/LinZZGLM17} worked on fairness-aware group recommendation, and designed a multi-objective optimization model to minimize the utility gap between group members.
Burke \etal~\cite{burke2017balanced} proposed a regularization-based matrix completion method to reweigh different neighbors, in order to balance the fairness between protected and unprotected neighbors in collaborative recommendation.
Zhu \etal~\cite{FATR} presented a fairness-aware tensor-based recommendation approach, which uses sensitive latent factor matrix to isolate sensitive features and then uses a regularizer to extract sensitive information which taints other factors.
More recently, going beyond the pointwise fairness metrics in ranking, Beutel \etal~\cite{DBLP:conf/kdd/BeutelCDQWWHZHC19} considered pairwise fairness of user preference between clicked and unclicked items, and offered a new regularizer to encourage improving this metric.


Besides in optimization objective, regularization also has been added in the ranking policy to address the unfairness issue. \cite{DBLP:conf/sigir/MorikSHJ20} considers the problem in dynamic ranking system, where the ranking function dynamically evolves based on the feedback that users provide, and present a new sorting criterion FairCo as follows:
\begin{align}
\quad \sigma_{\tau}=\underset{i \in \mathcal{I}}{\operatorname{argsort}}\left(\hat{R}(i \mid u)+\lambda \operatorname{err}_{\tau}(i)\right)
\end{align}
where the error term $\operatorname{err}$ measures the fairness violation has been introduced. The intuition behind FairCo is that the error term pushes the items from the underexposed groups upwards in the ranking lists. Fairness-aware constraints have also been introduced in learning-to-rank (LTR) models \cite{singh2019policy,yadav2021policy}. Efficient policy-gradient algorithms have been developed for model optimization \cite{DBLP:conf/sigir/Oosterhuis21,yadav2021policy}.

\textbf{Adversarial Learning.}
Similar with the idea of LFR (\cf \eqref{equ:fairness-lfr}), the line of adversarial learning aims to get fairness as a side-effect of fair representation.
The basic idea is to play a min-max game between the prediction model and an adversary model, where the adversary tries to predict the sensitive attributes from the data representations, so minimizing the performance of the adversary is to remove the information pertinent to the sensitive attributes in the representation.
Towards this goal, a general framework, \emph{Adversarial Learned Fair Representation (ALFR)}, is proposed in~\cite{ALFR} which is formulated as follows:
\begin{gather}
    \max_{\phi}\min_{\theta}\Lapl=\alpha C_{\theta}(X,R) + \beta D_{\theta,\phi}(R,A) + \gamma E_{\theta}(Y,R)
\end{gather}
where $C_{\theta}(\cdot)$ is the reconstruction loss to quantify the information retained in the representations $R$ about the data $X$ by the ability of an encoder or decoder network; $E_{\theta}(\cdot)$ is to predict $Y$ from $R$ via a predictor network; $\theta$ encompasses the parameters of the encoder/decoder and predictor networks; and $D_{\theta,\phi}(\cdot)$ is to quantify the independence between the representation $R$ and the sensitive attributes $A$ via an adversary network: $R\rightarrow A$. Assuming $A$ is binary, $D_{\theta,\phi}(\cdot)$ is formulated as log-loss for binary adversary network $f$:
\begin{gather}
    D= \Space{E}_{X,A} A\cdot\log(f(R))+(1-A)\cdot\log(1-f(R))
\end{gather}
which satisfies the fairness constraint of demographic parity.
Maximizing $D_{\theta,\phi}(\cdot)$ is to optimize the adversary's parameters $\phi$, while minimizing $D_{\theta,\phi}(\cdot)$ is to optimize the representation parameters $\theta$.

Only recently have researchers considered this line in the field of recommendation.
For example, Zhu \etal~ \cite{zhu2020measuring} leveraged adversarial learning to enhance the score distribution similarity between different groups.   Bose~\etal~\cite{CompositionalFairness} and Wu \etal~ \cite{wu2021learning} extended the ALFR framework by enforcing compositional fairness constraints on graph embeddings for multiple sensitive attributes, which are evaluated in the scenarios of item or friendship recommendation. Wherein, instead of fair \wrt single sensitive attribute, it makes the graph embeddings be invariant \wrt different combinations of sensitive attributes by employing a compositional encoder in the adversary network.
Building upon the ALFR framework, Beigi~\etal~\cite{DBLP:conf/wsdm/BeigiMGAN020} proposed a framework termed \emph{recommendation with attribute protection (RAP)} to recommend items based on user preference, while simultaneously defensing against private-attribute inference attacks.
In particular, the prediction and adversarial networks are instantiated as the sensitive attribute inference attacker and the Bayesian personalized recommender, respectively. Analogically, Li \etal~ \cite{DBLP:conf/sigir/LiCXGZ21} targeted at personalized counterfactual fairness with leveraging adversarial learning to isolated the personalized sensitive attributes.

\begin{figure}[t!]
    \centering
    \includegraphics[width=0.49\textwidth]{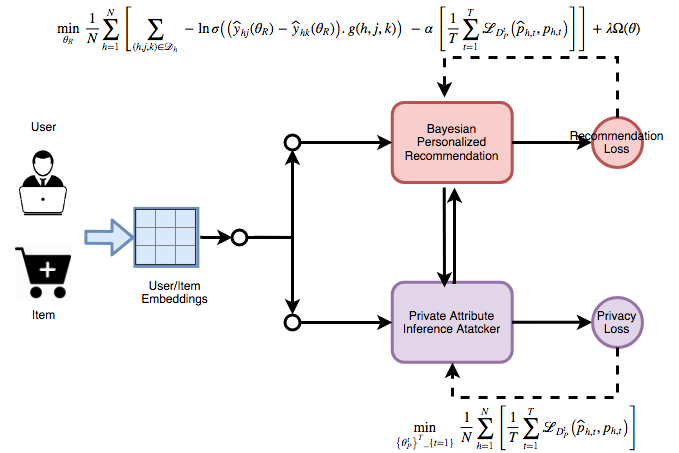}
     \caption{Recommendation with attribute protection (RAP)~\cite{DBLP:conf/wsdm/BeigiMGAN020}. The figure was reproduced from the Arxiv\protect\footnotemark~with authors' permission.}
    \label{fg:rap}
\end{figure}
\footnotetext{\url{https://arxiv.org/pdf/1911.09872.pdf}}

\textbf{Causal Modeling.}
Inspired by the success of causal modeling~\cite{pearl2009causality}, studying fairness from the causal perspective~\cite{DBLP:conf/aaai/ZhangB18,DBLP:conf/aaai/NabiS18,DBLP:conf/kdd/WuZW18,Counterfactual-Fairness,wu2019counterfactual} has attracted increasing attentions.
In general, fairness is formulated as the causal effect of the sensitive attribute, which is evaluated by applying counterfactual interventions over a causal graph.
For example, Wu~\etal~\cite{DBLP:conf/kdd/WuZW18} focused on fairness-aware ranking, and argued that the fairness constraints based on statistical parity hardly measure the discriminatory effect.
Hence, they built a causal graph that consists of the discrete profile attributes and the continuous score, and proposed a path-specific effect technique to detect and remove both direct and indirect rank bias.
Kusner~\etal~\cite{Counterfactual-Fairness} introduced the notion of counterfactual fairness, which is derived from Pearl's causal model~\cite{pearl2009causality}.
It considers the causal effect by evaluating the counterfactual intervention --- more formally, for a particular individual, whether its prediction in the real world is identical to that in the counterfactual world where the individual's sensitive attributes had been different.

\textbf{Others.} There are some other strategies on unfairness. For example, Li \etal~ \cite{li2021leave} proposed to add an autoencoder layer when learning user and item representation. This treatment can enforce that the specific unique properties of all users and items are sufficiently preserved in the representation, mitigating the bias towards mainstream users. Islam \etal~ \cite{islam2021debiasing} first computed a group-specific bias direction, and then debiased each user representation by subtracting its component in that direction. Ge \etal~ \cite{ge2021towards} studied on the problem of long-term fairness and proposed a fairness-constrained reinforcement learning algorithm to adapt dynamic fairness requirement.

\subsection{Methods for Mitigating Loop Effect}

Practise recommender systems usually create a pernicious feedback loop, which will create bias and further intensify bias over time. To deal with this problem, besides the aforementioned strategies on a specific bias, a surge of methods have been proposed recently to reduce the iterated bias that occurs during the successive interaction between users and recommender system.

\textbf{Uniform data.}   Leveraging uniform data is the most straightforward way to address the problem. To collect uniform data, this kind of methods intervene in the system by using a random logging policy instead of a normal recommendation policy. That is, for each user, they do not use the recommendation model for item delivery, but instead randomly select some items and rank them with a uniform distribution \cite{liu2020general,jiang2019degenerate}. The uniform data often provide gold-standard unbiased information because it breaks the feedback loop and is not affected by various biases. However, the uniform policy would inevitably hurt users' experience and the revenue of the platform, thus it is usually restricted to a small percentage of online traffic. Therefore, how to correct the bias with a small uniform data is a key research question.    Yuan \etal \cite{yuan2019improving} learn a imputation model from the uniform data and apply the model to impute the labels of all displayed or non-displayed items. Rosenfeld \etal~ \cite{rosenfeld2017predicting} and Bonner \etal \cite{bonner2018causal} employ two recommendation models for the biased data and uniform data, and further use a regularization term to transfer the knowledge between the models; Liu \etal~ \cite{liu2020general} leverage knowledge distillation to extract information from uniform data to learn a unbiased recommendation model. Yu \etal~ \cite{yu2020influence} leverage influence function to reweight training instances so that it has less loss in an unbiased validation set. Chen \etal~ \cite{chen2021autodebias} proposed to learn the optimal debiasing configures from the unbiased data.

\begin{figure}[t!]
\centering
\includegraphics[width=0.42\textwidth]{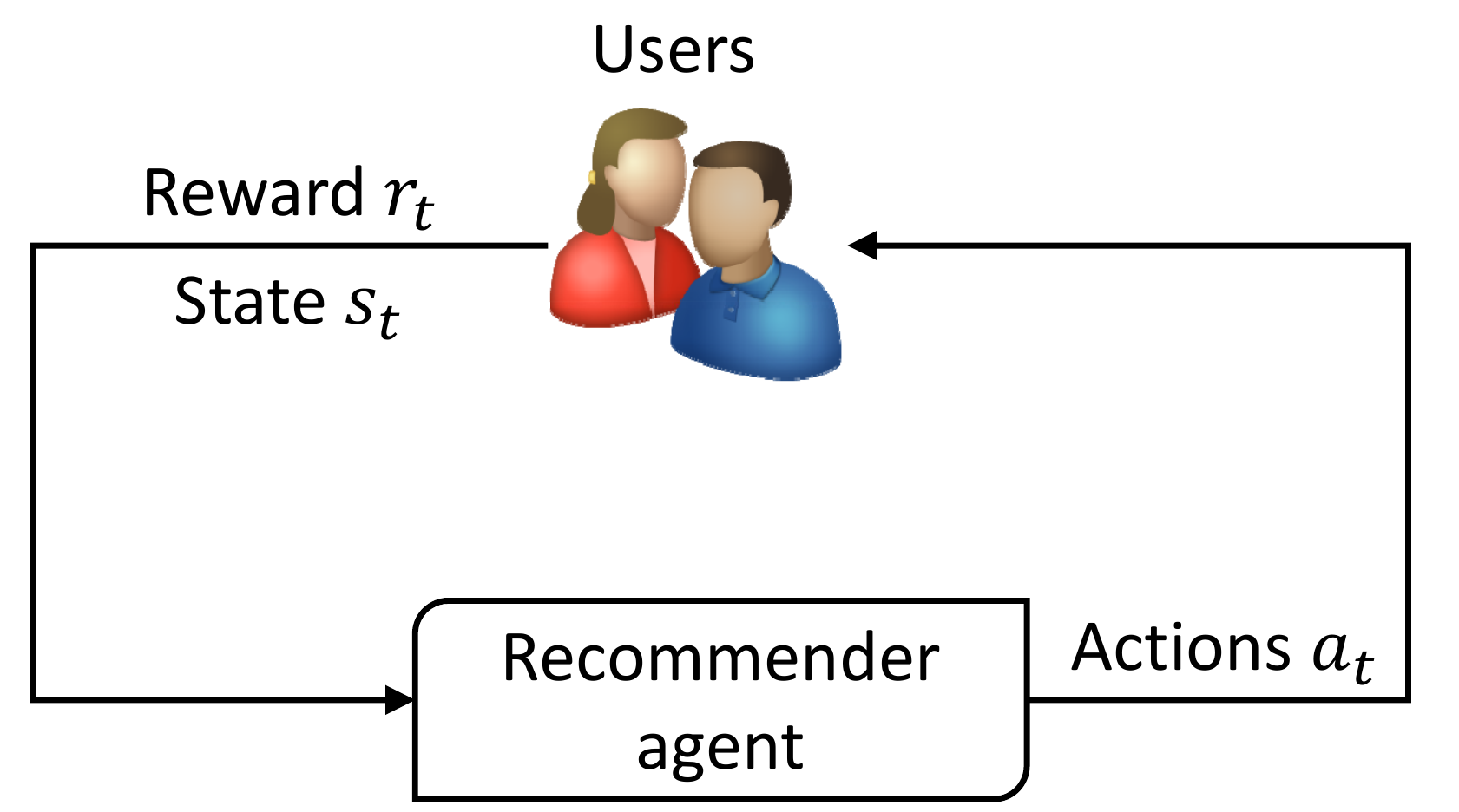}
 \caption{The system-user interactions with a RL aggent. The figure is plotted referring to \cite{zhao2018recommendations} with permission. }
\label{fg:rl}
\end{figure}

\textbf{Reinforcement learning.} Collecting uniform data with a random policy is not a satisfactory strategy as it hurts recommendation performance. Smarter recommendation strategy or policy needs to be explored. There exists an exploration-exploitation dilemma in recommender system, where the exploitation is to recommend items that are predicted to best match users' preference, while the exploration is to recommend items randomly to collect more unbiased user feedback to better capture user preference. To deal with this problem, a large number of work explores interactive recommendation by building a reinforcement learning (RL) agent. Figure \ref{fg:rl} illustrates the system-user interactions with a RL agent. Different from traditional recommendation methods, RL considers the information seeking tasks as sequential interactions between an RL agent (system) and users (environment). During the interaction, the agent can continuously update its strategies $\pi$ according to users' history information or feedback (i.e. state $s_t$) and generates a list of items (i.e. action $a_t$) that best match users' preferences or explore users' preference for long term reward. Then, the users will give the feedback (i.e. rewards $r_t$, such as ratings or clicks) on the recommendation lists to update the agent. Therefore, RL could balance the competition between the exploitation and exploration and maximize each user's long term satisfaction with the system \cite{zhao2019deep}. Some recent work \cite{li2010contextual,wang2017interactive,wang2017factorization,zhao2013interactive} balance exploitation and exploration in bandit setting with $\varepsilon$-greedy, Boltzmann Exploration or Upper Confidence Bounds (UCB). Some work estimates action-value reward function $Q(s,a)$ with Q network using the Bellman equation and finds the best strategy with the largest function value \cite{zhao2020jointly,chen2018stabilizing,zhao2018recommendations,zheng2018drn}. Also, the actor network has been adopted recently to learn the best policy by maximizing the long term reward \cite{chen2019large,zhao2018deep,zhao2017deep,wang2018reinforcement}.

A challenge of RL-based recommender is how to evaluate a policy. It is best to deploy it online, e.g., in the form of an A/B test, which however is expensive and time-consuming in terms of engineering and logistic overhead and also may harm the user experience when the policy is not mature \cite{jagerman2019people}. Off-policy evaluation is an alternative strategy that uses historical interaction data to estimate the performance of a new policy. However, off-policy evaluation will suffer from bias as the data are collected by an existing biased logging policy instead of uniform policy. To correct the data bias, Chen \etal~ \cite{chen2019top} proposes to weight the policy gradient with the inverse of the probability of historical policy. Inspired by \cite{chen2019top},  some work \cite{jagerman2019people,mcinerney2020counterfactual,swaminathan2017off} further explore off-policy evaluation for non-stationary recommendation environments or slate recommendation. However, as claimed by Jeunen \etal \cite{jeunen2020joint}, existing off-policy learning methods usually fail due to stochastic and sparse rewards. Therefore, they \cite{jeunen2020joint} further propose to leverage supervised signal with IPS strategy to better evaluate a policy. Nevertheless, off-policy evaluation is still a challenging task especially when the historical policy is not provided, which deserves for further exploration.

\textbf{Others.}  There are some other strategies to mitigate the loop effect. Sun \etal \cite{sun2019debiasing} leverage blind spot term to let items be close to each other in the latent space. Sinha \etal \cite{sinha2016deconvolving} provide an algorithm for deconvolving feedback loops to recover users' truth rating values.

\section{Future Work}

This section discusses open issues and point out some future directions.

\subsection{Evaluation of Propensity Scores}
As mentioned before, Inverse Propensity Score is a conventional strategy to debias. However, the effectiveness and unbiasedness of an IPS strategy are guaranteed only when the propensity scores are properly specified. How to obtain proper propensity scores remains an important research question. Existing methods usually assume the ideal propensities are given. Although the evaluation of propensity scores in some simple scenarios, e.g. for position bias, have been explored, evaluating propensity scores in more complex scenarios, such as for selection bias or exposure bias, is still an open problem and deserves further exploration.

\subsection{General Debiasing Framework}
From former studies, we can find that existing methods are usually designed for just addressing one or two specific biases. However, in the real world, various biases usually occur simultaneously. For example, users usually rate the items that they like and their rating values are influenced by the public opinions, where conformity bias and selection bias are mixed in the collected data. Besides, the distribution of rated user-item pairs is usually inclined to popular items or specific users groups, making the recommendation results easily suffer from popularity bias and unfairness. It is imperative that recommender systems require a general debiasing framework to handle the mixture of biases. It is a promising but largely under-explored area where more studies are expected. Although challenging, the simple case --- the mixture of just two or three biases --- is worth to be explored first.

IPS or its variants, which have been successfully applied for various biases, are a promising solution for this problem. It will be interesting and valuable to explore a novel IPS-based framework, which summarizes the applications of IPS on different kinds of biases and provides a general propensity score learning algorithm.

\subsection{Better Evaluation}
How to evaluate a recommender system in an unbiased manner? It is an essential question for both researchers and practitioners in this area. Existing methods either require accurate propensity scores or rely on a considerable amount of unbiased data. However, the accuracy of the former can not be guaranteed, while the latter hurts user experience and is usually constrained on a very small percentage of online traffic. Uniform data provides gold-standard unbiased information but its small scale makes it insufficient to thoroughly evaluate a recommendation model due to high variance. Exploring new evaluators using large-scale biased data and small-size unbiased data will be an interesting direction. More theoretical studies are expected, analyzing the expectation, bounds and confidences of the proposed evaluator.

Due to popularity bias and unfairness, the evaluation exhibits more difficulties. Different work usually adopts different evaluation criteria of popularity bias or unfairness. This creates an inconsistent reporting of scores, with each author reporting their own assortment of results. The performance or comparisons of existing methods can not be well understood. As such, we believe that a suite of benchmark datasets and standard evaluations metrics should be proposed.

\subsection{Knowlege-enhanced Debiasing}
It is natural that exploiting the abundant auxiliary information would improve the efficacy of debiasing. Recent years have seen some examples that leverage attributes of users or items to correct biases in recommendation. An interesting direction is how to better exploit this auxiliary information as the attributes are not isolated but connected with each other forming a knowledge graph. The knowledge graph captures much more rich information, which could be useful to understand the data bias. For example, given a user $u_1$ watches movies $i_1$ and $i_2$, both of which are directed by the same person $p_1$ and of the same genre $p_2$. From the knowledge graph, we can deduce that the $u_1$ are highly likely to have known the movies that connect with entities $i_1$, $i_2$, $p_1$ or $p_2$. This exposure information is important for exposure bias correction. Another advantage of knowledge graph is its generality. All data, data sources, and databases of every type can be represented and operationalized by the knowledge graph. Knowledge graph would be a powerful tool for developing a feature-enhanced general debiasing framework.

\subsection{Explanation and Reasoning with Causal Graph}
Cause graph is an effective mathematical tool for elucidating potentially causal relationships from data, deriving causal relationships from combinations of knowledge and data, predicting the effects of actions, and evaluating explanations for observed events and scenarios. As such, it is highly promising for the debiasing tasks in recommendation. On the one hand, the key of debaising is to reason the occurrence, cause, and effect over recommendation models or data. Most biases can be understood with mild cause assumptions and additional confounding factors in the causal graph. The effect of bias also can be inferred through the casual paths in the graph. On the other hand, recommendation is usually considered as an intervention analogous to treating a patient with a specific drug, where counterfactual reasoning needs to be conducted. What happens if the recommended items are exposed to the users? Causal graph provides potentials to answer this question. The formulated unbiased recommendation criteria can be derived with causal graph.

Nowadays, making explainable recommendations is increasingly important as it helps to improve the transparency, persuasiveness, effectiveness, trustworthiness, and satisfaction of a RS. Explainable recommendation and debiasing are highly related in the sense that they both address the problem of why: they both need to answer why certain items are recommended by the algorithm. When causal graph is promising to address the bias problem in a RS, it can also provide opportunities to give explanation from the strong causal paths in the graph.

To this end, the next step would to be to design better and suitable causal graph, which is capable of reasoning, debiasing,  and explanation. We believe causal model will bring the recommendation research into a new frontier.

\subsection{Dynamic Bias}
In real world, biases are usually dynamic rather than static. For example, the fashion of clothes changes frequently; users experience many new items and may get new friends every days; the recommendation system will update its recommendation strategy periodically; etc. All in all, factors or biases often evolve with the time going by. It will be interesting and valuable to explore how bias evolves and analyze how the dynamic bias affects a RS.

\subsection{Double-edged Sword of Bias}
Bias is not always harmful. For example, popularity bias has been validated as a double-edged sword \cite{zhao2021popularity,zhang2021causal}. Popularity bias not only results from conformity but also item quality. Appropriately leveraging popularity bias in recommendation may improve the performance. It will be interesting and valuable to explore the double-edged nature of other biases, fostering their benign effects while circumventing their harmful.

\subsection{Fairness-Accuracy Trade-off}
The trade-off between accuracy and fairness is of importance in recommendation scenarios, where equally treating different groups \wrt sensitive attributes has been shown to sacrifice the recommendation performance.
Hence, it inspires us to (1) identify specific unfairness issues; and (2) define the fairness criteria carefully to cover a wide range of use cases; and 3) design some controllable methods, where the trade-off between fairness and accuracy can be controlled.
Moreover, existing methods largely assume that the sensitive attributes of users (or items, groups) are provided as part of the input.
Such assumptions might not hold in certain real-world scenarios --- for example, in collaborative filtering, user profiles including sensitive attributes like age and gender cause different patterns of their behaviors; however, such profiles are unobserved but implicitly affect the recommendation performance.
A research direction is to understand the dimensions of causality and design fairness-aware collaborative filtering algorithms in case sensitive attributes are not readily available.

\section{Conclusions}
In this article, with reviewing more than 180 papers, we systematically summarize the seven kinds of biases in recommendation, along with providing their definitions and characteristics. We further devise a taxonomy to organize and position existing debiasing approaches, with discussing their strengths and weaknesses. We list some open problems and research topics worth to be further explored. We hope this survey can benefit the researchers and practitioners who are keen to understand the biases in recommendation and inspire more research work in this area.

\bibliographystyle{ACM-Reference-Format}
\bibliography{sigproc}
\end{document}